\documentclass[twocolumn,
showpacs,
showkeys,
preprintnumbers,
nofootinbib,
superscriptaddress,
amsmath,
amssymb,
floatfix,
secnumarabic,
aps,
pra,
a4paper,
notitlepage,
]{revtex4}%

\usepackage[colorlinks=true,urlcolor=blue]{hyperref}
\usepackage{graphicx}
\usepackage{epsfig}

\newcommand{\beq}{\begin{equation}}
\newcommand{\eeq}{\end{equation}}
\newcommand{\beqar}{\begin{eqnarray}}
\newcommand{\eeqar}{\end{eqnarray}}

\def\dalam{\hbox
{\vrule\vbox{\hrule\hbox to 1ex{ \hfill}\kern 1 ex\hrule}\vrule}}
\def\sign{\hbox{sign}}
\def\1/2{\hbox{$ {1 \over 2}$ }}

\def\h{\hbar}
\def\i/h{{i \over \h}}

\def\ch{\cosh}
\def\sh{\sinh}

\def\a{\alpha}

 \def\D{\Delta}
\def\l{\lambda} 
\def\e{\epsilon}

\def\s{\sigma}

\def\m{\mu}
\def\n{\nu}

\def\k{\kappa}

\def\<{\langle}
\def\>{\rangle}

\def\({\left(}
\def\[{\left[}
\def\){\right)}
\def\]{\right]}

\usepackage{subfigure}

\usepackage{tikz}
\usetikzlibrary{decorations.pathreplacing}
\usetikzlibrary{decorations.pathmorphing}    %Для рисования контуров интегрирования 2
\usepackage{multirow}
\usepackage{dcolumn}		%выравнивание по точке в таблицах, использовать d вместо c в спецификации колонки
\newcolumntype{.}{D{.}{.}{-1}}
\newcolumntype{i}[1]{D{.}{.}{#1}}

\newcommand{\mc}[1]{\multicolumn{1}{c}{#1}} % handy shortcut macro

\newcommand{\hm}[1]{#1\nobreak\discretionary{}{\hbox{\ensuremath{#1}}}{}}

\newcommand{\myfrac}[2]{{\ifmmode{}^{#1}\!/_{\!#2}\else${}^{#1}\!/_{\!#2}$\fi}}

\newcommand{\minitab}[2][c]{\begin{tabular}{#1}#2\end{tabular}}

\begin{document}
\sloppy

\title{Estimating the radiative part of QED effects in superheavy nuclear quasimolecules}

\author{A.~Roenko}
\email{roenko@physics.msu.ru} \affiliation{Department of Physics, Moscow State
University, 119991, Leninsky Gory, Moscow, Russia}

%%remember to change the email and affiliation!!!!!
\author{K.~Sveshnikov}
\email{costa@bog.msu.ru} \affiliation{Department of Physics, Moscow State
University, 119991, Leninsky Gory, Moscow, Russia}

%Remember to change the date
\date{\today}

%%%%%%%%%%%%%%%%%%%

\begin{abstract}

 A method for calculating the electronic levels in the compact superheavy nuclear quasi-molecules, based on  solving  the two-center Dirac equation using the multipole expansion of  two-center potential, is developed. For the internuclear distances up to $d\sim100$~fm such technique reveals a quite fast convergence and allows for computing the electronic levels in such  systems with accuracy  $\sim 10^{-6} $.
The critical distances $R_{cr}$ between the nuclei for $1\sigma_g$ and $1\sigma_u$ electronic levels  in the region $Z\simeq 87-100$ are calculated.
By means of the same technique  the  shifts of electronic levels due to the effective interaction $\Delta U_{AMM}$ of the electron's magnetic anomaly with the Coulomb field of the closely spaced heavy nuclei are evaluated as a function of the internuclear distance and the charge of the nuclei,  non-perturbatively both in $Z\a$ and (partially) in $\a/\pi$. It is shown, that the levels shifts near the lower continuum decrease with the enlarging size of the system of Coulomb sources both in the absolute units and in units of $Z^4 \a^5 / \pi n^3$. The last result is generalized to the whole self-energy contribution to the level shifts and so to the possible behavior of radiative part of QED-effects with virtual photon exchange  near  the lower continuum in the overcritical region.

\end{abstract}

%%pacs  numbers should be changed - see http://publish.aps.org/PACS
%\pacs{12.20.Ds, 31.15.A-, 31.30.J-, 34.10.+x}

\maketitle
%\tableofcontents
%%%%%%%%%%%%%%%%%%%%

%%%%%%%%%%%%%%%%%%%%%%%
\section{Introduction}\label{sec:intro}
In view of the  planned experiments on heavy ions collisions at FAIR (Darmstadt) and NICA (Dubna), the study  of electronic levels and QED-corrections in the compact nuclear quasi-molecules with large $Z$ turns out to be of special interest. The supercritical region, when the total charge of the colliding nuclei exceeds $Z_{cr} \simeq 170$, deserves a separate attention, since in this case QED predicts the non-perturbative vacuum reconstruction, which should be followed by a series of non-trivial effects, including the vacuum positron emission~\cite[and refs. therein]{Greiner1985a, Ruffini2010,  Rafelski2016}. However, the long-term experiments at GSI (Darmstadt) and Argonne National Lab didn't succeed in the  unambiguous conclusion of the status of the overcritical region, what promotes the question of the possible role of nonlinearity in the QED-effects for $Z > Z_{cr}$ to be quite actual~\cite{Rafelski2016, Ruffini2010, Schwerdtfeger2015, Sveshnikov2013, Davydov2017, Voronina2017}. In particular, the recent essentially non-perturbative results for the vacuum polarization energy for $Z > Z_{cr}$ confirm that in the supercritical region the behavior of the QED-effects could be substantially different from the perturbative case~\cite{Davydov2017, Voronina2017, Sveshnikov2017}. At the same time, the completely non-perturbative in $Z\a$ and (partially) in $\a/\pi$ evaluation of  level shifts near the threshold of the lower continuum in the superheavy H-like atoms with $Z\a >1$, caused  by the interaction $\D U_{AMM}$ of the  electron's magnetic anomaly (AMM)   with the  Coulomb field of the atomic nucleus by taking into account  its dynamical screening at small distances $ \ll 1/m$, has shown that the growth rate of the contribution from $\D U_{AMM}$ reaches its maximum at $ Z \sim 140-150$, while by  further increase into the supercritical region $Z\gg Z_{cr}$ the shift of levels  near  the lower continuum decreases monotonically to zero in agreement with perturbative calculations~\cite{Roenko2017,Roenko2017a,Roenko2017b}.

 A closely related and even more actual problem is the magnitude  of  radiative QED-effects in the low-energy heavy ions collisions. As long as the distance between nuclei is about the atomic scale, the QED-corrections to the electronic levels are well described by the perturbation theory (PT)~\cite[and refs. therein]{Korobov2008, Komasa2011, Liu2014}. However,  as soon as the nuclei approach (adiabatically) slowly each other,  a transition to the supercritical region could occur, where the validity of PT is questionable~\cite{Davydov2017, Voronina2017, Sveshnikov2017}. Therefore, the investigation of those separate QED-effects, which allow for an essentially  non-perturbative analysis, turns out to be  quite important. In particular, such an effect is the interaction $\D U_{AMM}$ of the electron's AMM with the Coulomb field of external sources  with large $Z$, that for a single superheavy nucleus has been considered in detail in~\cite{Roenko2017,Roenko2017a,Roenko2017b}.

  Because the electronic AMM is a specific radiative effect, rather than  an immanent  property of the electron,  for strong external fields or  extremely small distances $\ll 1/m$ the dependence of the electronic form factor $F_2(q^2)$ on the momentum transfer should be taken into account from the very beginning~\cite{Lautrup1976, Barut1977, Geiger1988, Roenko2017,Roenko2017a,Roenko2017b}. In the general case the calculation of the  form factors, responsible for AMM, should be implemented via  self-consistent treatment of both    the external field  and the  electronic wave function (WF)~\cite{ Barut1977}. However, for the stationary electronic states even in superheavy atoms or in low-energy heavy ions collisions  the mean radius of the electronic WF substantially exceeds the size of the nuclear cluster, and so  the correct estimate for the corresponding  form factors could be made within  PT in $\a/\pi$.   Since the one-loop correction to the vertex function   can be represented via electronic form factors $F_1(q^2)$ and $F_2(q^2)$ in the form~\cite{Itzykson1980}
\begin{equation}\label{eq:02}
\Gamma^\mu (q^2) = \gamma^\mu F_1(q^2) + \frac{i}{2m} F_2(q^2) \sigma^{\mu\nu} q_\nu \, ,
\end{equation}
for strong  fields or  extremely small distances $\ll 1/m$  the  Dirac-Pauli term in the Dirac equation (DE) should be replaced by the expression
\begin{equation}\label{eq:03}
\D U_{AMM}(\vec{r}\,)= \frac{e}{ 2m}\, \s^{\m\n}\partial_\mu \mathcal{A}^{(cl)}_\nu(\vec{r}\,) ,
\end{equation}
where
\begin{equation}\label{eq:04}
\mathcal{A}^{(cl)}_\mu(\vec{r}\,) = \frac{1}{(2\pi)^3} \int \! d\vec{q} \ e^{i \vec{q}\,\vec{r}}\, \tilde{A}_\mu^{(cl)}(\vec{q}\,) F_2(-\vec{q}\,^2) \, ,
\end{equation}
and $\tilde{A}_\mu^{(cl)}(\vec{q}\,)$ is the Fourier-transform of the classical external field $A^{(cl)}_\mu(\vec{r}\,)$.  The detailed non-perturbative analysis of the interaction~\eqref{eq:03} between the electronic AMM and the Coulomb field of the superheavy nucleus  with $Z \a > 1$ shows~\cite{Roenko2017,Roenko2017a,Roenko2017b} that the growth rate of the contribution from $\D U_{AMM}$ reveals a significantly non-monotonic behavior with increasing $Z$. And since  $\D U_{AMM}$ is a part of the self-energy contribution to the total radiative shift of levels, the investigation of this effective interaction should be very useful for estimating the possible behavior of radiative QED-effects with virtual photon exchange in the overcritical region.
%For the H-like atom with nuclear charge $Z_{cr}=170$ calculations, performed in Refs.~\cite{Cheng1976, Soff1982} to the leading order in $\a/\pi$ with complete dependence on $Z\a$, give  $\D E_{SE}(1s_{1/2}) \hm\simeq 11.0 \text{\ keV} $, while the contribution from $\D U_{AMM}$ is of the order of 1~keV~\cite{Roenko2017}. And although contribution from $\D U_{AMM}$ is not a dominant radiative correction to the binding energy of the ground state in superheavy atoms, for a number of electronic levels in H-like atom it have the same power-low dependence on $Z$ as the full self-energy contribution~\cite{Roenko2017a}.

In particular, it is shown in this paper that for rather small internuclear distances $\D U_{AMM}$ can be quite effectively treated at the same footing with high accuracy calculations of electronic levels themselves within the  multipole expansion without spoiling the convergence of the latter. Thus, it allows to compute nonperturbatively the shifts of the electronic levels caused by $\D U_{AMM}$ both in $Z\a$ and (partially) in $\alpha/\pi$ (since $\alpha/\pi$ enters as a factor in the coupling constant for $\D U_{AMM}$), and thereby to find out the dependence of this QED-effect on the distance between nuclei and their charges.  Besides this, the significantly non-monotonic behavior of the growth rate, namely, the surplus over $Z^4$  for an H-like atom, when the electronic levels approach the threshold of the lower continuum~\cite{Roenko2017}, is shown to be typical only for a very compact nuclear quasi-molecule. With the increasing distance between nuclei the behavior of the growth rate of the contribution from $\D U_{AMM}$ becomes smoother, although the discrete levels continue to dive into the lower continuum as before, while the levels shift near the lower threshold  decreases monotonically in units of $Z^4 \a^5 / \pi n^3$.

%%%%%%%%%%%%%%%%%%%%%%%%%%%%%%
\section{The two-center DE with $\Delta U_{AMM}$}\label{sec:eq} % in spherical coordinates}\label{sec:eq}

\subsection{Methods for dealing with two-center DE}

In contrast to the case of a single nucleus, the inclusion of the effective interaction $\D U_{AMM}$ into the two-center DE requires the development of a special technique.
The problem of finding the electronic levels in two-nuclei quasi-molecules occurs also in the calculation of such an important parameter of two colliding nuclei as the critical distance $R_{cr}$ between nuclei, for which the binding energy of the electronic ground state amounts to two electron rest masses.
In general,
the determination of the critical distance requires for solving the two-center DE, what in the case of the low-energy ions collision  can be carried out in the adiabatic approximation, for which a large number of different  methods has been developed.   These are  the methods, based on the linear combination of atomic orbitals (LCAO) and variational ones~\cite{Mueller1973, Tupitsyn2010, Gail2003, Toshima1988, Momberger1993, Fillion-Gourdeau2012}, as well as various methods of numerical integration of DE using the finite elements techniques and lattice calculations~\cite{Sundholm1994, Kullie2001, Busic2004, Yang1991} (the most comprehensive review of the latter is given in Refs.~\cite{Artemyev2010, Tupitsyn2010, Tupitsyn2014}). The most recent evaluations~\cite{Popov2001, Mironova2015} give for the critical distance $R_{cr}$ in the symmetric quasi-molecules with total charge $Z_\Sigma \sim 170-190$ the result $\sim 10-50$~fm, which turns out to be of the same order as  the aggregated diameter of colliding nuclei. Therefore, the most of methods, based on LCAO and widely used in quantum chemistry, are not applicable in this case, since with the decreasing quasi-molecular size they require for a substantial increase of the number of basis elements. The calculations of $R_{cr}$ by various methods, using an expansion of the electronic WF in a finite set of basis functions combined with the variational principle, are considered in Refs.~\cite{Mueller1973, Lisin1980, Lisin1977, Matveev2000, Popov2001, Tupitsyn2010, Mironova2015}.%, Rafelski1976}.

At the same time, the Coulomb field of two closely spaced nuclei differs from the spherically symmetric one only slightly, what motivates to solve DE directly in the spherical coordinates, associated with the center of mass of the quasi-molecule, by means of the multipole expansion of the potential. 
%The validity of this approach is confirmed by the results of calculation of the critical distances $R_{cr}$ in the monopole approximation~\cite{Soff1979, Wietschorke1979}, that turn out to be  quite close to the values obtained by other methods. 
The validity of this approach was demonstrated in Refs.~\cite{Rafelski1976,Rafelski1976a,Soff1979}, where the binding energy calculations for point-like nuclei have been carried out using the multipole expansion. For very closely spaced extended nuclei the monopole approximation turns out to be enough for calculation the critical distances $R_{cr}$ with an accuracy of about 5\%~\cite{Soff1979, Wietschorke1979}. %, that confirmed by comparison with the values obtained by other methods. 
Moreover, the monopole approximation can be used by computing the parameters of the resonances, arising by diving of discrete electronic levels  into the lower continuum~\cite{Ackad2007, Ackad2008}. However, when the  internuclear distance increases, the monopole approximation becomes too rough, and so the higher multipoles of expansion of the two-center Coulomb potential  are required. Accounting for the higher multipoles is crucial in evaluating the ionisation probability of electronic shells in the heavy ions collisions~\cite{McConnell2012, McConnell2013, Bondarev2015}. And although the calculations of  $R_{cr}$ using multipole moments up to $l_{max}=4$ refine the results of the monopole approximation, this multipole truncation is not enough for the most heavy nuclei in the region $Z\sim 88-100$~\cite{Marsman2011}.

In this paper we present the results of numerical solving the stationary two-center DE in the spherical coordinates by means of the multipole expansion of both the Coulomb potential and the effective interaction due to electronic AMM~\eqref{eq:03} combined with the computer algebra tools, which provide to evaluate analytically all the Coulomb multipole moments in the model of nuclear charge as a uniformly charged ball. On the example of the quasi-molecule U$_2^{183+}$, the energy of the lowest electronic levels is explored as a function of the distance $d$ between the nuclei  and truncations in the electronic WF  expansion $\kappa_{max}$ and in the multipole expansion of two-center potential $l_{max}$ (up to $\k_{max}\sim 50$ and $l_{max}\sim 100$). It turns out that for the compact nuclear quasi-molecules ($d\lesssim 100$~fm) the suggested method reveals a quite fast convergence in $l_{max}$, $\kappa_{max}$, that allows one to compute the electronic levels in such a system with an accuracy  of $10^{-6} \sim 10^{-7}$. Within the approach developed the critical distances between heavy nuclei with charge $Z\sim 87-100$ for electronic levels $1\sigma_g$ and $1\sigma_u$ are calculated. The obtained values $R_{cr}$  coincide well with other results~\cite{Lisin1980, Tupitsyn2010, Marsman2011, Mironova2015} and significantly improve the results of the monopole approximation~\cite{Soff1979, Wietschorke1979}.

\subsection{Two-center DE with $\Delta U_{AMM}$ in the multipole expansion}

Let us consider the simplest nuclear quasi-molecule, which consists of two identical nuclei with charge $Z$, spaced by the distance $d=2a$. The  reference frame is chosen in such a way that the centers of the nuclei are placed on the $z$-axis with coordinates $(0,0,\pm a)$. The  external field for the electron in this case is given by
\begin{equation}\label{eq:04b}
A^{(cl)}_\mu(\vec{r}\,)\hm=\delta_{0,\mu} \(\Phi_0(|\vec{r}-\vec{a}|)+\Phi_0(|\vec{r}+\vec{a}|)\),
\end{equation}
where $\vec{a}=a \, \vec{e_z}$, while $\Phi_0(r)$ is the spherically-symmetric Coulomb field of a single nucleus, which is defined in the usual way through the nuclear charge distribution $ \rho_0(r) $, specified later.

Taking into account that to the leading order $F_2(0) \equiv {\Delta g_{free}}/{2} \simeq {\a}/{2\pi} $, upon substitution the Fourier-transform of~\eqref{eq:04b} into~\eqref{eq:04} and angular integration, one obtains
\begin{gather}\label{eq:05}
\mathcal{A}^{(cl)}_\mu(r) = -\frac{\Delta g_{free}}{2}\,\frac{e}{4\pi}\, V(\vec{r}\,) \, \delta_{\mu,0} \ , \\ V(\vec{r}\,)=Z\(\frac{ c(|\vec{r}-\vec{a}\,|)}{|\vec{r}-\vec{a}\,|}+\frac{c(|\vec{r}+\vec{a}\,|)}{|\vec{r}+\vec{a}\,|}\)  ,
\end{gather}
where
\begin{equation}\label{eq:05b}
c(r)= 2 \int\limits_0^\infty \! q dq \ \sin q r \(-\frac{1}{Z e} \,\tilde{\Phi}_0(q) \) \frac{1}{\pi} \frac{F_2(-q^2)}{F_2(0)} \, ,
\end{equation}
while $\tilde{\Phi}_0(q)$ is the Fourier-transform of the potential $\Phi_0(r)$. %The expression $\Delta g_{free}\,c(r)$ could be interpreted as the dependence of the electronic AMM on the distance from the nucleus center~\cite{Roenko2017}.

In the next step, the effective potential~\eqref{eq:03} should be rewritten as the following commutator
\begin{equation}\label{eq:06}
\D U_{AMM}(\vec{r}\,)= -\lambda \,[ \vec{\gamma} \cdot \vec{p}\,,\,V(\vec{r}\,)] \, ,
\end{equation}
where $\lambda=\alpha^2/4\pi m$, $\a = e^2/4\pi$. %Далее получим общий вид системы уравнений для нахождения электронных уровней, после чего конкретизируем модель распределения объёмной плотности заряда в ядре и функцию $c(r)$ для соответствующего кулоновского потенциала~$\Phi_0(r)$.
So the general form of DE for an electron with account for the additional effective interaction due to AMM~\eqref{eq:06} takes the form ($\hbar=c=m=1$)
\begin{equation} \label{eq:1}
\left( \vec{\alpha}\vec{p} + \beta + W (\vec{r}\,) + \Delta U_{AMM}(\vec{r}\,) \right) \psi = \epsilon \psi \, ,
\end{equation} %$-\alpha \, U(\vec{r}\,) = e A^{(cl)}_0(\vec{r}\,)$
where $W(\vec{r}\,)$ is the Coulomb interaction of the electron with the nuclei. For our purposes it is convenient to present $W(\vec r\,)$  in the form $W(\vec{r}\,) = -\alpha \, U(\vec{r}\,)$, where
\begin{equation} \label{eq:2}
U(\vec{r}\,) = \int d\vec{r}\,'\ \frac{\rho(\vec{r}\,)}{|\vec{r}-\vec{r}\,'|}\, \, ,
\end{equation}
while $\rho(\vec{r}\,)\hm = \rho_0(\vec{r}-\vec{a}\,) + \rho_0(\vec{r}+\vec{a}\,)$.

From the Eq.~\eqref{eq:1} for the upper $i\varphi$ and the lower $\chi$ components of the Dirac bispinor there follows
\begin{equation}\label{eq:6}
\begin{split}
i \left( \vec{\sigma}\vec{p} + \lambda \left[ \vec{\sigma} \vec{p} \,, V (\vec{r}\,) \right] \right) \varphi & = \left( \epsilon + 1 + \alpha \, U(\vec{r}\,) \right) \chi \, ,\\
i \left( \vec{\sigma}\vec{p} - \lambda \left[ \vec{\sigma} \vec{p} \,, V (\vec{r}\,) \right] \right) \chi & = - \left( \epsilon - 1 + \alpha \, U(\vec{r}\,) \right) \varphi  \, ,
\end{split}
\end{equation}

 Since the considered system possesses axial symmetry, the projection of the total momentum of the electron on $z$-axis is conserved. Moreover, for such a choice of reference frame $\rho(\vec{r}\,) \hm =\rho(-\vec{r}\,)$, hence, the electronic levels can be classified by parity. The spinors $\varphi$, $\chi$, corresponding to the solution of  Eqs.~\eqref{eq:6} with definite value of $m_j$, are seeded now as the following expansions in spherical spinors
\begin{equation}\label{eq:7}
\varphi = \sum_{\kappa = \pm 1}^{\pm N}  f_\k\, X_{\k, m_j}\, , \qquad \chi = \sum_{\kappa = \pm 1}^{\pm N}  g_\k\, X_{-\k, m_j}\, ,
\end{equation}
where $\k=\pm(j+\myfrac{1}{2})$,  the notations $X_{-|\k|, m_j} \equiv\Omega_{jlm_j}$ and $X_{|\k|, m_j} \hm \equiv (\vec{\sigma} \vec{n} )\, \Omega_{jlm_j}$ are used (the definition of spherical harmonics and spinors follows Ref.~\cite{Bateman1953}), while the radial functions $f_\k$, $g_\k$ can be taken real. The index $\k$ in the expansions~\eqref{eq:7} for an even case takes the values $\k=-1,+2,-3,+4,\dots$, while for odd one $\k=+1,-2,+3,-4,\dots$.

As a result, for the energy levels one obtains the spectral problem in the form of the following system of equations for the radial functions $f_\k$, $g_\k$
\begin{equation}\label{eq:8}
\begin{split}
\partial_r f_{\k} &+ \frac{1+\k}{r}f_{\k}  + \lambda \sum_{\bar{\k}} M_{\k;\bar{\k}}(r)\, f_{\bar{\kappa}} = \\
  &\hspace{4em} = (1+\epsilon)g_{\k} + \a \sum_{\bar{\k}} N_{-\k;-\bar{\k}}(r)\, g_{\bar{\kappa}}\, ,\\
\partial_r g_{\k} &+ \frac{1-\k}{r}g_{\k}  - \lambda \sum_{\bar{\k}} M_{-\k;-\bar{\k}}(r)\, g_{\bar{\kappa}} = \\
  &\hspace{4em} = (1-\epsilon)f_{\k} - \a \sum_{\bar{\k}} N_{\k;\bar{\k}}(r)\, f_{\bar{\kappa}} \, ,
\end{split}
\end{equation}
where the coefficient functions $N_{\k;\bar{\k}}(r)$, $M_{\k;\bar{\k}}(r)$ are expressed via matrix elements of the potential $U(\vec{r}\,)$ and of the commutator $\left[ \vec{\sigma} \vec{p} \,, V (\vec{r}\,) \right]$ over spherical spinors

\begin{equation}\begin{split}\label{eq:9}
%\begin{align}\label{eq:9}
N_{\k;\bar{\k}}(r)&\equiv \langle X_{\k, m_j} | U (\vec{r}\,) | X_{\bar{\k}, m_j} \rangle, \\ M_{\k;\bar{\k}}(r) &\equiv i \langle X_{-\k, m_j} | [\, \vec{\sigma} \vec{p} \,, V(\vec{r}\,) ] | X_{\bar{\k}, m_j} \rangle .
\end{split} \end{equation}
%\end{align}
%\end{subequations}
%%%%%%%%%%%%%%%%%

 To find the matrix elements $N_{\k;\bar{\k}}(r)$, $M_{\k;\bar{\k}}(r)$, the multipole expansions of the axially symmetric potentials $U(\vec{r}\,)$ and $V(\vec{r}\,)$ are used, what permits to separate the  radial and angular variables
\begin{equation}\label{eq:11}
\begin{split}
U(\vec{r}\,)= \sum_{n} U_{n}(r) P_n(\cos \vartheta)\, , \\ V(\vec{r}\,)= \sum_{n} V_n(r) P_n(\cos \vartheta) \, .
\end{split}
\end{equation}
%%%%%%%%%%%%%
In~\eqref{eq:11} the multipole moments $U_{n}(r)$ contain the complete dependence on the nuclear charge density
%%%%%%%%%%%%%%%% one col
%\begin{equation}\label{eq:12}
%U_{n}(r) = \int \! d\vec{r}\,' \rho(\vec{r}\,') \(\Theta(r-r') \, \frac{r'^{n}}{r^{n+1}} + \Theta(r'-r) \, \frac{r^{n}}{r'^{n+1}} \) P_n(\cos \vartheta')\, ,% ,\quad F_n(r,r')=\Theta(r-r') \, \frac{r'^{n}}{r^{n+1}} + \Theta(r'-r) \, \frac{r^{n}}{r'^{n+1}} \, .
%\end{equation}
%%%%%%%%%%%%%%% two col
\begin{multline}\label{eq:12}
U_{n}(r) = \int \! d\vec{r}\,' \rho(\vec{r}\,') \, P_n(\cos \vartheta') \, \times \\ \times \Big(\Theta(r-r') \, \frac{r'^{n}}{r^{n+1}} +  \Theta(r'-r) \, \frac{r^{n}}{r'^{n+1}} \Big) \, ,
\end{multline}
%%%%%%%%%%%%%%
while the general form of the multipoles $V_n(r)$ is the following
\begin{equation}\label{eq:14}
V_n(r)=\frac{2n+1}{2} \int\limits_{0}^{\pi} \!\sin \theta  \,d\theta \, P_n(\cos \theta) V(\vec{r}\, )\, .
\end{equation}

In the next step, the matrix elements $M_{\k;\bar{\k}}(r)$ can be reduced to the form $\langle X_{\k, m_j} | V(\vec{r}\,) | X_{\bar{\k}, m_j} \rangle$, whence  for the coefficient functions~\eqref{eq:9} one obtains the following final expressions
\begin{equation}\label{eq:15a}
\begin{split}
N_{\k;\bar{\k}}(r)  &= \sum_{n} U_{n}(r)\, W^{\varsigma}_{\bar{\varsigma}} (n;l_\k;l_{\bar{\k}}), \\
M_{\k;\bar{\k}}(r)  &= \sum_{n} \left( \partial_r + \frac{\k-\bar{\k}}{r} \right) V_n(r)\, W^{\varsigma}_{\bar{\varsigma}}(n;l_\k;l_{\bar{\k}}),
\end{split}
\end{equation}
where $$\varsigma=\sign(-\k)=\left\{ \begin{matrix} -, &\ \k>0 \ , \\ +, &\ \k<0 \ , \end{matrix} \right. \hspace{1.5em} l_\k=\left\{ \begin{matrix} \k,\phantom{|-1|} &\ \k>0 \ , \\ |\k|-1, &\ \k<0 \ , \end{matrix} \right. $$
while the factors $$W^{\varsigma}_{\bar{\varsigma}}(n;l_\k;l_{\bar{\k}})\equiv \langle X_{\k, m_j} | P_n(\cos \vartheta) | X_{\bar{\k}, m_j} \rangle $$ for the fixed sign of $\kappa$ and $\bar{\kappa}$ are given by the next combinations of the $3j$-symbols:

\begin{equation*} \begin{split}
	W^+_\pm(n;l;s)=&\sqrt{(l+m_j+\myfrac{1}{2})(s\pm m_j+\myfrac{1}{2})} \ w_n^-(l;s) \pm  \\
	 & \pm \sqrt{(l-m_j+\myfrac{1}{2})(s\mp m_j+\myfrac{1}{2})} \ w_n^+(l;s)\, ,\\
	W^-_\pm(n;l;s)=&\sqrt{(l-m_j+\myfrac{1}{2})(s\pm m_j+\myfrac{1}{2})} \ w_n^-(l;s) \mp  \\
	 & \pm \sqrt{(l+m_j+\myfrac{1}{2})(s\mp m_j+\myfrac{1}{2})} \ w_n^+(l;s)\, ,\\
\end{split}
\end{equation*}
%%%%%%%%%%%%%%%%
where%\vspace{-1em}
%%%%%%%%%%%%%%% one col
%\begin{equation}
%w_n^\pm(l;s)=(-1)^{m_j\pm \myfrac{1}{2}}  \begin{pmatrix} l & n & s \\ -(m_j\pm \myfrac{1}{2}) & 0 & m_j \pm \myfrac{1}{2} \end{pmatrix} \begin{pmatrix} l & n & s \\ 0 & 0 & 0 \end{pmatrix} \ .
%\end{equation}
%%%%%%%%%%%%%%%% two col
\begin{multline*}
w_n^\pm(l;s)=\\=(-1)^{m_j\pm \myfrac{1}{2}}  \begin{pmatrix} l & n & s \\ -(m_j\pm \myfrac{1}{2}) & 0 & m_j \pm \myfrac{1}{2} \end{pmatrix} \begin{pmatrix} l & n & s \\ 0 & 0 & 0 \end{pmatrix}.
\end{multline*}
%%%%%%%%%%%%%%%%%%%

\begin{subequations}
For the levels of definite parity the expansions in spherical spinors~\eqref{eq:7} and the system of equations~\eqref{eq:8} could be written as follows. One should pass everywhere from the summation over $\k=\pm1,\dots,\pm N$  to summation over $k=0,1,\dots,\tilde{N}$, $\tilde{N}=N/2-1$, thence each series in $\k$ splits into two series in $k$, since $\k$ is defined via different expressions in terms of $k$ for the cases  $\k \gtrless 0$. The expansion~\eqref{eq:7} contains the same number of terms with positive and negative $\k$, provided the cutoff number $N$ is even. For the even level everywhere in Eqs.~\eqref{eq:7},~\eqref{eq:8} it is necessary to perform the substitution~\eqref{eq:7evv}, while for odd one~---~\eqref{eq:7odd}:
\begin{align}
&\k \rightarrow -2k-1\, ,\quad \k<0\, ; &\k \rightarrow 2k+2\, ,\quad \k>0\, , \label{eq:7evv} \\
&\k \rightarrow -2k-2\, ,\quad\k<0\, ;  &\k \rightarrow 2k+1\, ,\quad \k>0\, . \label{eq:7odd}
\end{align}
\end{subequations}
The systems of equations for the levels of definite parity are given in the explicit form in the Appendix~\ref{sec:app-syst}.

Within the described method of finding the electronic levels in two-center quasi-molecules both the point-like  and the extended nuclei with arbitrary charge density $\rho_0(r)$ could be considered. In the subsequent sections we specify the nuclear charge density, calculate the corresponding multipole moments~\eqref{eq:12},~\eqref{eq:14} and present the results of computation the energy of the electronic levels $1\sigma_q$ and $1\sigma_u$. It should be noted that in the case of $\lambda=0$ the Eqs.~\eqref{eq:8} describe the purely Coulomb case, when the effective interaction $\Delta U_{AMM}$ is absent, and so the coefficient functions $M_{\k,\bar{\k}}(r)$ do not enter the equations.

In addition let us mention that  for the Dirac-Pauli term (i.e., in the approximation $F_2(q^2)\hm\simeq F_2(0)$) the multipole moments $V_n$ coincide with $U_n$. Indeed, since in this case the integral in~\eqref{eq:05b} results to $c(r) = r\,\Phi_0(r)$, it turns out that $V(\vec{r}\,)=U(\vec{r}\,)$, and the expression~\eqref{eq:14} precisely  coincides with~\eqref{eq:12}.

%%%%%%%%%%%%%%%%%%%%%%%%%%%%%%%%
\section{Calculation of the multipole moments $U_n$, $V_n$}\label{sec:formfact}

%%%%%%%%%%%%%%%%%%%%%%%%%%%%%%%%
\subsection{The point-like nuclei}
For a symmetric two-nuclei quasi-molecule the charge density $\rho(\vec{r}\,)$ and the function  $V(\vec{r}\,)$ are even, hence, the multipole moments~\eqref{eq:12},~\eqref{eq:14} do not vanish for even $n$ only due to the symmetric properties of the Legendre polynomials. In the case of a point-like nuclei $\rho_0(\vec{r}\,)\hm = Z \, \delta(\vec{r}\,)$, and so the multipoles~\eqref{eq:12} take the simplest form
\begin{equation}\label{eq:16}
U_{n}(r)= 2 Z \(\Theta(r-a) \, \frac{a^{n}}{r^{n+1}} + \Theta(a-r) \, \frac{r^{n}}{a^{n+1}} \).
\end{equation}

The function $c(r)$, that modulates the behavior of the electronic AMM at small distances from the Coulomb source, for the point-like nucleus is given by~\cite{Lautrup1976, Roenko2017}
\begin{equation}\label{eq:16a}
c(r)=1- \int\limits_{4m^2}^\infty \! \frac{dQ^2}{Q^2} \, e^{- Q r}\, \frac{1}{\pi}\, \frac{\text{Im}\, F_2(Q^2)}{F_2(0)} \, ,
\end{equation}
where for the one-loop form factor in~\eqref{eq:02} one gets~\cite{Barbieri1972} $$ \frac{1}{\pi}\, \text{Im}\, F_2(Q^2) = 2 F_2(0)\, \dfrac{m^2}{Q^2} \, \dfrac{1}{\sqrt{1-4m^2/Q^2}}.$$
Upon taking into account both the expansion
\begin{multline*}
\frac{e^{ik|\vec{r}-\vec{a}|}}{|\vec{r}-\vec{a}|} = \\ =\frac{i \pi}{2 \sqrt{ra}}\sum_{l=0}^\infty (2l+1) J_{l+1/2}(k r_<) H^{(1)}_{l+1/2} (kr_>) P_l(\cos \theta)\, ,
\end{multline*}
%%%%%%%%%%%%%%%%%%
where $r_<=\min(r,a)$, $r_>=\max(r,a)$, and the orthogonality properties of the Legendre polynomials, the final answer for the multipole moments~\eqref{eq:14} of the point-like nuclei for even $n$ could be written in the form
%%%%%%%%%%%%%%%% one col
%\begin{equation}\label{eq:16c}
%V_n(r)=2Z\left(\frac{r_<^n}{r_>^{n+1}}-\frac{2n+1}{2\sqrt{rd}}\int\limits_{4m^2}^\infty \!\frac{dQ^2}{Q^2}\, \frac{\text{Im}\, F_2(Q^2)}{F_2(0)}\, i\,J_{n+1/2}(iQr_<) \,H^{(1)}_{n+1/2}(iQr_>)\right)\, , \quad n=\text{even}\, .
%\end{equation}
%%%%%%%%%%%%%%%%%% two col
\begin{multline}\label{eq:16c}
V_n(r)=2Z\Bigg(\frac{r_<^n}{r_>^{n+1}}-\frac{2n+1}{2\sqrt{ra}}\int\limits_{4m^2}^\infty \!\frac{dQ^2}{Q^2}\, \frac{\text{Im}\, F_2(Q^2)}{F_2(0)}\, \times \\ \times i\,J_{n+1/2}(iQr_<) \,H^{(1)}_{n+1/2}(iQr_>)\Bigg), \ n=\text{even}\, .
\end{multline}
%%%%%%%%%%%%%%%

%%%%%%%%%%%%%%%%%%%%%%%%%%%%%%%%
\subsection{The extended nuclei}
The extended nuclei are treated as uniformly charged balls with the radius $R$, that is defined via $Z$ by means of the (simplified) expression $R=1.228935 \, (2.5 Z)^\myfrac{1}{3}$~fm, as in Ref.~\cite{Roenko2017}. In this case, as it will be shown below, the multipole moments~\eqref{eq:12} and the coefficient functions $N_{\k;\bar{\k}}(r)$~\eqref{eq:9} can be evaluated analytically, what significantly simplifies the further calculations. The charge density of a nucleus in this model is defined as  $\rho_0(\vec{r}\,) \hm = \Theta(R-r)\,{3Z}/{(4\pi R^3)}$, while the multipoles $U_{n}$ for even $n$ are given by the following expression

\begin{multline}\label{eq:17}
U_{n}(r) = \frac{6 Z}{2R^3}\int\limits_{a-R}^{a+R}\! r'^2 dr' \, \zeta_{n}(r')\, \times \\ \times \Big(\Theta(r-r') \, \frac{r'^{n}}{r^{n+1}} + \Theta(r'-r) \, \frac{r^{n}}{r'^{n+1}} \Big),
\end{multline}
%%%%%%%%%%%%%%%%%%%%%
where
%%%%%%%%%%%%%%%%%% one col
%\begin{equation}
%\zeta_{n}(r') = \int\limits_{-1}^1 dx' P_{n}(x') \Theta\left(R - \sqrt{d^2-r'^2-2dr'x'}\right)  = \dfrac{P_{n-1}(x_0')-P_{n+1}(x_0')}{2n+1}
%\end{equation}
%%%%%%%%%%%%%%%% two col
\begin{multline}
\zeta_{n}(r') = \int\limits_{-1}^1 dx' P_{n}(x') \Theta\left(R - \sqrt{a^2-r'^2-2ar'x'}\right)  = \\ =\dfrac{P_{n-1}(x_0')-P_{n+1}(x_0')}{2n+1}
\end{multline}
%%%%%%%%%%%%%%%%%%%%
and $x_0'={(a^2+r'^2-R^2)}/{(2ar')}$. Since the  the following integrals~\eqref{eq:19} can be easily calculated analytically for any $n$:
\begin{equation}\label{eq:19}\begin{split}
\int\limits_{a-R}^{a+R} dr' r'^{n+2}\zeta_n(r') = \frac{2 a^n R ^3}{3} \, , \\  \int\limits_{a-R}^{a+R} dr' r'^{-n+1} \zeta_n(r') = \frac{2 R ^3}{3 a^{n+1}}\ ,
\end{split}\end{equation}
%%%%%%%%%%%%%%%%%%%
as a result, one obtains
%%%%%%%%%%%%%%%%%%%%%% one col
%\begin{equation}\label{eq:20}
%U_{n}(r) = \left\{
%\begin{aligned}
%&\dfrac{2Z d^{n}}{r^{n+1}} \,, &r < d-R \, ,\\
%&\dfrac{3 Z}{R^3}\bigg(\frac{1}{r^{n+1}}\int\limits_{d-R}^r \! dr'\, {r'}^{n+2} \zeta_{n}(r') +
%r^{n}\int\limits_r^{d+R} \! dr'\,\frac{1}{{r'}^{n-1}} \,\zeta_{n}(r')\bigg), & |r-d| \leq R \, ,\\
%&\dfrac{2Z r^{n}}{d^{n+1}} \,, & r > d+R \, .\\
%\end{aligned}\right.
%\end{equation}
%%%%%%%%%%%%%%%%%%% two col
\begin{equation}\label{eq:20}
U_{n}(r) = \left\{
\begin{aligned}
&\dfrac{2Z a^{n}}{r^{n+1}} \,,\hspace{8.5em} r < a-R \, ,\\
&\dfrac{3 Z}{R^3}\bigg(\int\limits_{a-R}^r \! dr'\, \frac{{r'}^{n+2}}{r^{n+1}} \zeta_{n}(r') + {}\\
& \  +\!\! \int\limits_r^{a+R} \! ar'\,\frac{r^{n}}{{r'}^{n-1}} \,\zeta_{n}(r')\bigg),\ \ |r-a| \leq R \, ,\\
&\dfrac{2Z r^{n}}{a^{n+1}} \,, \hspace{8.5em} r > a+R \, .\\
\end{aligned}\right.
\end{equation}
%%%%%%%%%%%%%%%%%%%

It should be noted, that in the intermediate region $|r-a| \leq R$ it is also possible to calculate the integral over $r'$ in~\eqref{eq:20} analytically for any given $n$ by means of computer algebra, since the integrand turns out to be a rational function. However, unlike the integrals~\eqref{eq:19}, in the intermediate region $|r-a|\leq R$ the answer cannot be obtained in the general form for arbitrary $n$. The analytical expressions of the multipole moments~\eqref{eq:20} in the intermediate region are presented in the Appendix~\ref{sec:app-Fn} for a set of initial $n\, , \ 0\leq n \leq 12$.

For the  considered model of the extended nucleus with the radius $R$ the function $c(r)$ has been 
calculated in Ref.~\cite{Roenko2017} and equals to
\begin{subequations}\label{eq:20a}%
\begin{align}
c(r)&=1-\! \int\limits_{4m^2}^\infty \! \frac{dQ^2}{Q^2}\, \frac{3QR\ch QR - 3\sh QR}{R^3 Q^3} \times {}\nonumber\\
 & \hspace{5em} \times e^{- Q r}\, \frac{1}{\pi}\, \frac{\text{Im}\, F_2(Q^2)}{F_2(0)}, \quad r>R\, ,\\
c(r)&=\frac{(3 R^2 - r^2)}{2 R^3}\,r - \frac{r}{2 m^2 R^3} +\! \int\limits_{4m^2}^\infty \! \frac{dQ^2}{Q^2}\, \frac{3(QR+1)}{R^3 Q^3} \times {}\nonumber\\
 & \quad \ \times \sinh Qr \, e^{- Q R}\, \frac{1}{\pi}\, \frac{\text{Im}\, F_2(Q^2)}{F_2(0)} , \quad  r<R\, .
\end{align}
\end{subequations}
%%%%%%%%%%%%%%%%%%
The multipole moments $V_n$ are evaluated numerically using the expression~\eqref{eq:20a} directly from the definition~\eqref{eq:14}.

\begin{figure*}[tbh]
\subfigure{\label{pic:Fn1}
\includegraphics[width=.48\textwidth]{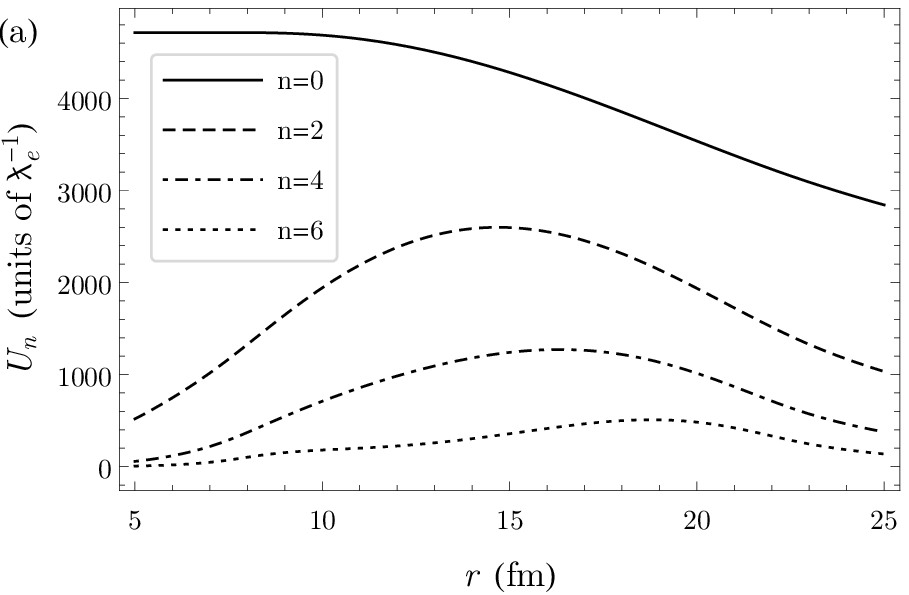} %Fn0-6+.eps}
}
\hfill
\subfigure{\label{pic:Fn2}
\includegraphics[width=.48\textwidth]{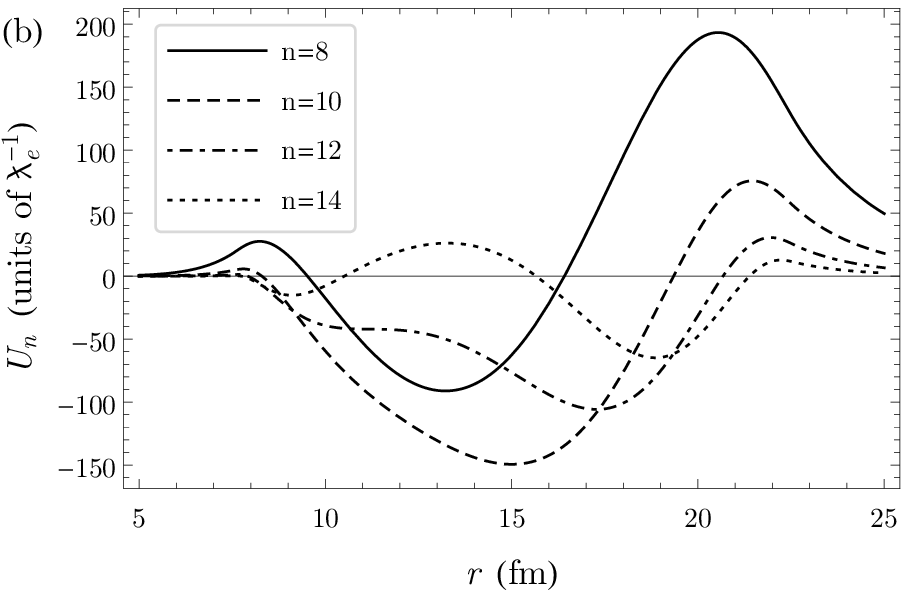} %Fn8-14+.eps}
}
\subfigure{\label{pic:Gn1}
\includegraphics[width=.48\textwidth]{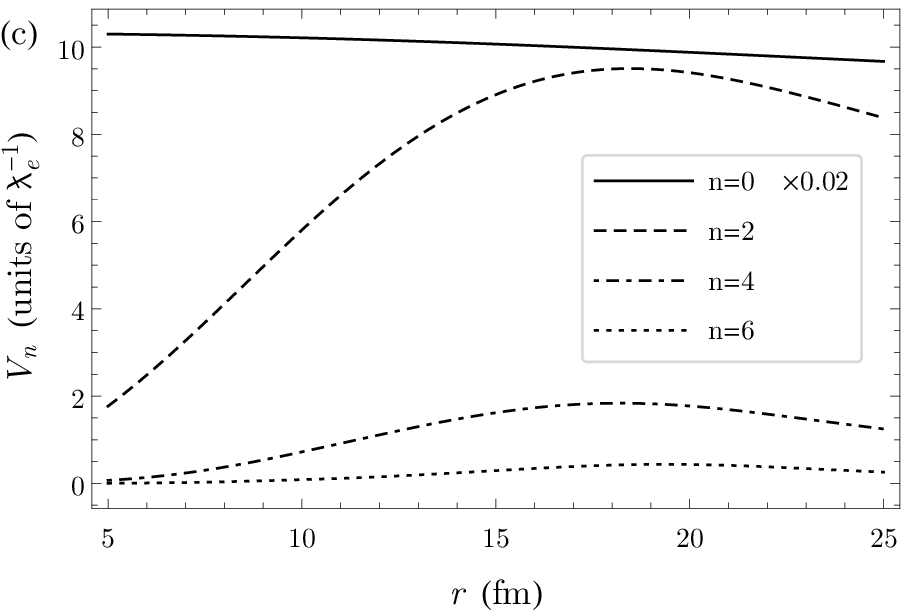} %Gn0-6+.eps}
}
\hfill
\subfigure{\label{pic:Gn2}
\includegraphics[width=.48\textwidth]{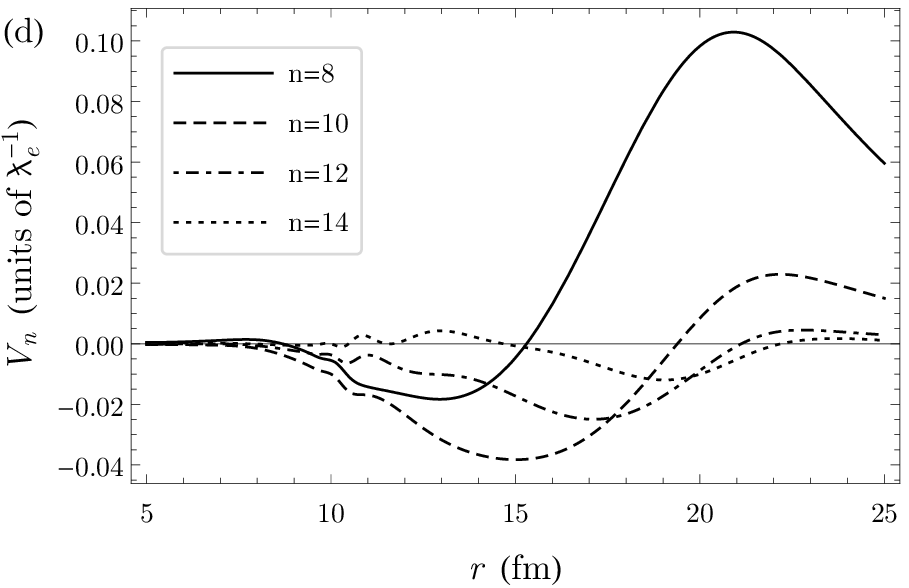} %Gn8-14+.eps}
}
 \caption{The multipoles $U_n$~\subref{pic:Fn1},~\subref{pic:Fn2} and $V_n$~\subref{pic:Gn1},~\subref{pic:Gn2} of the $\rm U-U$ system (the nuclear radius is $R\simeq 7.54$~fm) for the distances between the centers of the nuclei $d=30$~fm in the interval $(a-R, a+R)$; the multipoles of  order $n=0,2,4,6$ are shown  in Figs.~\subref{pic:Fn1},~\subref{pic:Gn1} (the multipole moment $V_0$ is scaled by a factor 1/50) and of order $n\hm=8,10,12,14$ --- in Figs.~\subref{pic:Fn2},\subref{pic:Gn2} (in units of inverse Compton wavelength for the electron $1/\lambdabar_e=mc/\hbar$).}\label{pic:Fn}
\end{figure*}

In the external regions $|r-a| > R$ the multipoles $U_n$ of the extended nuclei~\eqref{eq:20} reveal the same power-law behavior as the multipoles of the point-like nuclei~\eqref{eq:16}, but inside the nuclei their behavior turns out to be highly nontrivial. The behavior of the first multipoles $U_n$ of two Uranium nuclei separated by the distance $d=30$~fm in this region is shown in Figs.~\ref{pic:Fn1},~\ref{pic:Fn2}. Unlike the multipole moments of the point-like nuclei, which  possess a jump in the derivative at the point $r=a$, that is the sharper, the greater $n$, the multipoles~\eqref{eq:20} and their derivatives behave smoother, but from some $n$ (depending on $Z$ and $a$) begin to oscillate in the intermediate region. Thus, in the case of the extended nuclei  the coefficient functions $N_{\k;\bar{\k}}(r)$~\eqref{eq:9} turn out to be the smooth, in general  oscillating, rational functions.

%%%%%%%%%%%%%%%%%%%%%%%%%%%%%%%%%%%%%
%
%	Fig 1
%
%%%%%%%%%%%%%%%%%%%%%%%%%%%%%%%%%%%%%

The behavior of the multipoles $V_n$ in the same region is shown in Figs.~\ref{pic:Gn1},~\ref{pic:Gn2}. In comparison with the Coulomb multipoles $U_n$, the moments $V_n$ also oscillate in the intermediate region, but their magnitudes decrease much faster with growing $n$. So the inclusion of the high-order multipoles of the Coulomb potential is more important compared  to those from $\D U_{AMM}$. Consideration of the multipoles $U_n$, $V_n$ for other parameters $Z$ and $d$ leads to the same conclusion.

    The more detailed nuclear charge densities $\rho_0(\vec{r}\,)$ (for example, the Fermi distribution) could also be considered within our approach, but in these cases all the multipole moments~\eqref{eq:12} and~\eqref{eq:14} should be evaluated numerically. This circumstance insignificantly complicates the problem, but the consideration of the other models of charge distribution in the extended nuclei lies beyond the scope of this paper.

%%%%%%%%%%%%%%%%%%%%%%%%%%%%%%%%%%%
\section{Applicability and accuracy of the method}\label{sec:pre}
As it has been already discussed in Sec.~\ref{sec:intro}, the choice of the spherical coordinates is the more justified, the closer the symmetry of a system is to the spherical one, that means, the smaller the distance between nuclei. The whole set of the spherical spinors $X_{\k,m}$, which is used in the WF expansion~\eqref{eq:7}, forms a complete orthogonal basis, but for the numerical calculation one should truncate it. Thus, the number of harmonics $N$ used in the expansion~\eqref{eq:7}, which suffices to determine the electronic level with the given accuracy,  appears to be the natural parameter, characterizing applicability of the method. It should be noted that for the given cutoff $\kappa_{max}=N$ in the WF expansion~\eqref{eq:7}, the system of equations~\eqref{eq:8} includes the multipoles $U_n$, $V_n$ of the order $n\leq l_{max}=2N$, and all of them are involved in  further calculations. 

 In the first step, let us consider the pure Coulomb case (the Eqs.~\eqref{eq:8} with $\l=0$). The energies of the lowest even and odd electronic levels ($1\sigma_g$ and $1\sigma_u$) in the quasi-molecule $\rm U_2^{183+}$ are presented in Tabs.~\ref{tab:eNumA},~\ref{tab:eNumB} for  various truncations in the WF expansion~\eqref{eq:7}, including the monopole approximation, and for certain  distances between nuclei. In the case of $d=2a=38.5$~fm, which is the critical distance for the point-like Uranium nuclei, the values from the Ref.~\cite{Artemyev2010} are given for comparison, and the difference between the results for the extended nuclei (in the forth digit) originate from the different nuclear charge distributions. To achieve the accuracy $10^{-5}$ for the extended nuclei spaced by $d=38.5$~fm the cutoff $\k_{max}\sim12-14$ turns out to be enough. However, for the point-like nuclei the method converges too slowly, and the accuracy $10^{-4}$ is not achieved even with $\k_{max}\sim 40$, since the configuration is far from the spherical-symmetric one. The results of extrapolation to the region $\k_{max}\sim 50-200$ are also listed in Tab.~\ref{tab:eNumB}, that confirm the slow convergence of this technique for the point-like nuclei. When the extended nuclei move away from each other, the number of harmonics needed to achieve the specified accuracy increases (see Tab.~\ref{tab:eNumA}), but up to $d\simeq100$~fm it remains acceptable. It also follows from the presented data, that the monopole approximation shows an error about $\sim0.05\, mc^2$ in this region.
The accuracy of computing the energy of electronic levels does not change, when the effective potential $\D U_{AMM}$ is taken into account. Moreover, since the ratio $|V_n/V_0|$ decreases with the growing $n$ much faster, than $|U_n/U_0|$,  a fewer number of harmonics is needed to determine the levels shift due to $\Delta U_{AMM}$ with the given accuracy, than it is required to find the Coulomb  binding energy  (see Tab.~\ref{tab:eNumC}).

%%%%%%%%%%%%%%%%%%%%%%%%%%
%
%	Tab. 1
%
%						 	v2
%%%%%%%%%%%%%%%%%%%%%
\begin{table*}[tbh]
\center
 \caption{The electronic levels $1\sigma_g$ and $1\sigma_u$ in dependence on $\k_{max}=N$ in WF expansion~\eqref{eq:7} ($d=38.5$~fm, $d=77$~fm and $d=100$~fm $Z=92$, extended nuclei). The monopole approximation is denoted by $\k_{max}=1$. For comparison, the values from the Ref.~\cite{Artemyev2010} are given (in units of $mc^2$).}\label{tab:eNumA}
\begin{ruledtabular}
\begin{tabular}{i{3.0} *{6}{i{2.10}}}
%\toprule
 & \multicolumn{2}{c}{$d=38.5$~fm} & \multicolumn{2}{c}{$d=77$~fm} & \multicolumn{2}{c}{$d=100$~fm}  \\
 \cline{2-3}\cline{4-5}\cline{6-7}\noalign{\smallskip}
\mc{ $\k_{max}$}  & \mc{ $\e_{1\s_g}$ (extend)} & \mc{ $\e_{1\s_u}$ (extend)} & \mc{ $\e_{1\s_g}$ (extend) } & \mc{ $\e_{1\s_u}$ (extend)} & \mc{ $\e_{1\s_g}$ (extend) } & \mc{ $\e_{1\s_u}$ (extend) }\\
\colrule\noalign{\smallskip}
1 & {\bf -}{\bf 0}.{\bf 8}8068503 & {\bf -0}.{\bf 0}4750302 & {\bf -0}.{\bf 4}6966445 & {\bf 0}.{\bf 2}6473979 & {\bf -0}.{\bf 3}4300642 & {\bf 0}.{\bf 3}4750011 \\
%2 & {\bf -0}.{\bf }87412882 & {\bf -0}.{\bf }04140322 & {\bf -0}.{\bf }46522091 & {\bf 0}.{\bf 2}6864461 & {\bf -0}.{\bf 3}3943246 & {\bf 0}.{\bf 3}5053294 \\
4 & {\bf -0}.{\bf 92}053834 & {\bf -0}.{\bf 08}215949 & {\bf -0}.{\bf 5}0408585 & {\bf 0}.{\bf 2}3907531 & {\bf -0}.{\bf 3}7520439 & {\bf 0}.{\bf 3}2494307 \\
6 & {\bf -0}.{\bf 92}726296 & {\bf -0}.{\bf 08}813156 & {\bf -0}.{\bf 51}257777 & {\bf 0}.{\bf 23}251593 & {\bf -0}.{\bf 3}8349391 & {\bf 0}.{\bf 3}1892112 \\
8 & {\bf -0}.{\bf 928}46235 & {\bf -0}.{\bf 089}16654 & {\bf -0}.{\bf 51}541712 & {\bf 0}.{\bf 23}031937 & {\bf -0}.{\bf 38}651976 & {\bf 0}.{\bf 31}672205 \\
10 & {\bf -0}.{\bf 9286}5404 & {\bf -0}.{\bf 0893}6985 & {\bf -0}.{\bf 51}653404 & {\bf 0}.{\bf 22}945558 & {\bf -0}.{\bf 38}785684 & {\bf 0}.{\bf 31}575183 \\
12 & {\bf -0}.{\bf 92867}529 & {\bf -0}.{\bf 0893}8885 & {\bf -0}.{\bf 517}00152 & {\bf 0}.{\bf 229}09414 & {\bf -0}.{\bf 38}850518 & {\bf 0}.{\bf 31}528208 \\
14 & {\bf -0}.{\bf 92867}675 & {\bf -0}.{\bf 08939}018 & {\bf -0}.{\bf 517}19901 & {\bf 0}.{\bf 228}94143 & {\bf -0}.{\bf 388}83402 & {\bf 0}.{\bf 315}04405 \\
16 & {\bf -0}.{\bf 928677}46 & {\bf -0}.{\bf 08939}083 & {\bf -0}.{\bf 517}28014 & {\bf 0}.{\bf 2288}7865 & {\bf -0}.{\bf 389}00366 & {\bf 0}.{\bf 314}92131 \\
18 & {\bf -0}.{\bf 928677}86 & {\bf -0}.{\bf 089391}18 & {\bf -0}.{\bf 5173}1141 & {\bf 0}.{\bf 2288}5441 & {\bf -0}.{\bf 389}09093 & {\bf 0}.{\bf 314}85817 \\
20 & {\bf -0}.{\bf 928677}93 & {\bf -0}.{\bf 089391}25 & {\bf -0}.{\bf 51732}219 & {\bf 0}.{\bf 22884}603 & {\bf -0}.{\bf 3891}3499 & {\bf 0}.{\bf 3148}2628 \\
22 & {\bf -0}.{\bf 928677}95 & {\bf -0}.{\bf 089391}26 & {\bf -0}.{\bf 51732}521 & {\bf 0}.{\bf 22884}364 & {\bf -0}.{\bf 3891}5645 & {\bf 0}.{\bf 3148}1072 \\
24 & & & {\bf -0}.{\bf 517325}77 & {\bf 0}.{\bf 228843}24 & {\bf -0}.{\bf 3891}6634 & {\bf 0}.{\bf 3148}0354 \\
 26 & & & {\bf -0}.{\bf 517325}78 & {\bf 0}.{\bf 228843}21 & {\bf -0}.{\bf 38917}052 & {\bf 0}.{\bf 31480}049 \\
 28 & & & {\bf -0}.{\bf 517325}79 & {\bf 0}.{\bf 228843}20 & {\bf -0}.{\bf 38917}206 & {\bf 0}.{\bf 31479}937 \\
 30 & & & & & {\bf -0}.{\bf 389172}50 & {\bf 0}.{\bf 314799}04 \\
 32 & & & & & {\bf -0}.{\bf 389172}54 & {\bf 0}.{\bf 314799}00 \\
\colrule\noalign{\smallskip}
\mc{ Others~\cite{Artemyev2010}} & -0.92831 & -0.08908 & \mc{---} & \mc{---} & \mc{---} & \mc{---} \\
%\botrule
\end{tabular}
\end{ruledtabular}
\end{table*}
%%%%%%%%%%%%%%%%%%%%%

%%%%%%%%%%%%%%%%%%%%%%%%%%
%
%	Tab. 2
%
%%%%%%%%%%%%%%%%%%%%%%%%%%
%							v3
%%%%%%%%%%%%%%%%%%%%%%%%%%
\begin{table*}[tbh]
\center
\caption{The electronic levels $1\sigma_g$ and $1\sigma_u$ in dependence on $\k_{max}=N$ in WF expansion~\eqref{eq:7} ($d=38.5$~fm and $d=2/Z~\text{a.u.}\simeq 1150~\text{fm}$, $Z=92$). The monopole approximation is denoted by $\k_{max}=1$. For comparison, the values from the Ref.~\cite{Artemyev2010} are given  (in units of $mc^2$).}\label{tab:eNumB}\label{tab:eNum2Z}
%\footnotesize
\begin{ruledtabular}
\begin{tabular}{i{3.0} *{4}{i{2.10}}}
%\toprule
 & \multicolumn{2}{c}{$d=38.5$~fm} & \multicolumn{2}{c}{$d=2/Z\simeq 1150$~fm} \\
 \cline{2-3}\cline{4-5}\noalign{\smallskip}
\mc{ $\k_{max}$}  & \mc{ $\e_{1\s_g}$ (point-like) } & \mc{ $\e_{1\s_u}$ (point-like) } & \mc{$\e_{1\s_g}$ (extend)} & \mc{$\e_{1\s_g}$ (point-like)}\\
\colrule\noalign{\smallskip}
1 &  {\bf -0}.{\bf }90914697 & {\bf -0}.{\bf }07115473 & {\bf 0}.{\bf }53072585 & {\bf 0}.{\bf }53071572 \\
%2 & {\bf -0}.{\bf }87412882 & {\bf -0}.{\bf }04140322 & {\bf -0}.{\bf }90098022 & {\bf -0}.{\bf }06330628 \\
4 & {\bf -0}.{\bf 9}6368764 & {\bf -0}.{\bf 1}1864261 & {\bf 0}.{\bf 4}9199067 & {\bf 0}.{\bf 4}9196287 \\
%6 & {\bf -0}.{\bf 9}7967341 & {\bf -0}.{\bf 1}3289370\\
8 & {\bf -0}.{\bf 9}8633523 & {\bf -0}.{\bf 1}3883316 & {\bf 0}.{\bf 4}7813668 & {\bf 0}.{\bf 4}7808456 \\
%10 & {\bf -0}.{\bf 9}8983489 & {\bf -0}.{\bf 14}194896\\
12 & {\bf -0}.{\bf 99}193745 & {\bf -0}.{\bf 14}381799 & {\bf 0}.{\bf 4}7433742 & {\bf 0}.{\bf 4}7426618 \\
%14 & {\bf -0}.{\bf 99}331635 & {\bf -0}.{\bf 14}504198\\
16 & {\bf -0}.{\bf 99}427829 & {\bf -0}.{\bf 14}589479 & {\bf 0}.{\bf 47}268968 & {\bf 0}.{\bf 47}260245 \\
%18 & {\bf -0}.{\bf 99}498089 & {\bf -0}.{\bf 14}651701\\
20 & {\bf -0}.{\bf 99}551260 & {\bf -0}.{\bf 14}698746 & {\bf 0}.{\bf 47}180521 & {\bf 0}.{\bf 47}170405 \\
%22 & {\bf -0}.{\bf 99}592651 & {\bf -0}.{\bf 14}735339\\
24 & {\bf -0}.{\bf 99}625625 & {\bf -0}.{\bf 14}764470 & {\bf 0}.{\bf 47}126790 & {\bf 0}.{\bf 47}115429 \\
%26 & {\bf -0}.{\bf 99}652400 & {\bf -0}.{\bf 14}788110 \\
28 & {\bf -0}.{\bf 99}674496 & {\bf -0}.{\bf 14}807607 & {\bf 0}.{\bf 47}091394 & {\bf 0}.{\bf 47}078901 \\
%30 & {\bf -0}.{\bf 99}692984 & {\bf -0}.{\bf 14}823914 \\
32 & {\bf -0}.{\bf 99}708640 & {\bf -0}.{\bf 14}837716 & {\bf 0}.{\bf 47}066706 & {\bf 0}.{\bf 47}059171 \\
%34 & {\bf -0}.{\bf 99}722038 & {\bf -0}.{\bf 14}849522\\
36 & {\bf -0}.{\bf 99}733608 & {\bf -0}.{\bf 14}859715 & {\bf 0}.{\bf 47}048741 & {\bf 0}.{\bf 47}034236 \\
%38 & {\bf -0}.{\bf 99}743683 & {\bf -0}.{\bf 14}868588\\
40 & {\bf -0}.{\bf 99}752520 & {\bf -0}.{\bf 14}876368 & {\bf 0}.{\bf 47}035233 & {\bf 0}.{\bf 47}018184 \\
%42 & {\bf -0}.{\bf 99}760323 & {\bf -0}.{\bf 14}883235\\
44 & {\bf -0}.{\bf 99}767358 & {\bf -0}.{\bf 14}889334 & {\bf 0}.{\bf 47}024810 & {\bf 0}.{\bf 47}009245\\
48 & & & {\bf 0}.{\bf 47}016598 & {\bf 0}.{\bf 46}999902\\
\colrule\noalign{\smallskip}
\mc{ Extrap. }& & \\
50 & {\bf -0}.{\bf 99}7840(03) & {\bf -0}.{\bf 14}9040(66) & {\bf 0}.{\bf 47}0131(31) & {\bf 0}.{\bf 46}9956(69) \\
60 & {\bf -0}.{\bf 998}030(05) & {\bf -0}.{\bf 149}207(67) & {\bf 0}.{\bf 47}0002(53) & {\bf 0}.{\bf 46}9809(87) \\
80 & {\bf -0}.{\bf 998}24(252) & {\bf -0}.{\bf 149}39(420) & {\bf 0}.{\bf 469}86(785) & {\bf 0}.{\bf 469}64(458) \\
100 & {\bf -0}.{\bf 998}35(504) & {\bf -0}.{\bf 149}49(279) & {\bf 0}.{\bf 469}80(489) & {\bf 0}.{\bf 469}55(640) \\
140 & {\bf -0}.{\bf 998}4(6794) & {\bf -0}.{\bf 149}5(9134) & {\bf 0}.{\bf 469}7(5506) & {\bf 0}.{\bf 469}4(6723) \\
%%180 & -0}.{\bf 99852(2521616) & -0}.{\bf 14963(8593930) \\
200 & {\bf -0}.{\bf 998}5(4010) & {\bf -0}.{\bf 149}6(5401) & {\bf 0}.{\bf 469}7(3875) & {\bf 0}.{\bf 469}4(0958) \\
%%220 & -0}.{\bf 9985(53871515) & -0}.{\bf 1496(65467596) \\
%%260 & -0}.{\bf 9985(73913690) & -0}.{\bf 1496(82456069) \\
%%300 & {\bf -0}.{\bf 998}58(7697587) & {\bf -0}.{\bf 149}69(4002234) \\
\colrule\noalign{\smallskip}
\mc{ Others~\cite{Artemyev2010}} & -0.99842 & -0.14956 & 0.4697339 & 0.4693303 \\
%\botrule
\end{tabular}
\end{ruledtabular}
\end{table*}
%%%%%%%%%%%%%%%%%%%%%%%%%%%%%%%

%%%%%%%%%%%%%%%%%%%%%%%%%%
%
%	Tab. 3
%
%%%%%%%%%%%%%%%%%%%%%%%%%%
\begin{table*}[tbh]
\center
\caption{The shift of the levels $1\sigma_g$ and $1\sigma_u$ due to $\Delta U_{AMM}$ for a various total charge of nuclei $Z_{\Sigma}=2Z$ and internuclear distances $d$ in dependence on $\k_{max}=N$ in  WF expansion~\eqref{eq:7}  (in units of $mc^2$). The monopole approximation is denoted by $\k_{max}=1$.}\label{tab:eNumC}
\begin{ruledtabular}
\begin{tabular}{i{2.0} *{6}{i{2.10}}}
%\toprule
 & \multicolumn{2}{c}{$Z=88$,\ $d=20$~fm} & \multicolumn{2}{c}{$Z=92$,\ $d=38.5$~fm} & \multicolumn{2}{c}{$Z=92$,\ $d=77$~fm} \\
 \cline{2-3}\cline{4-5}\cline{6-7}\noalign{\smallskip}
\mc{ $\k_{max}$ } & \mc{ $\D\e_{1\s_g}$ } & \mc{ $\D\e_{1\s_u}$ } & \mc{ $\D\e_{1\s_g}$  } & \mc{ $\D\e_{1\s_u}$ } & \mc{ $\D\e_{1\s_g}$  } & \mc{ $\D\e_{1\s_u}$ } \\
\colrule\noalign{\smallskip}
  1 & {\bf  0}.{\bf 0020}2849 & {\bf  -0}.{\bf 0015}4317 & {\bf  0}.{\bf 0017}8063 & {\bf  -0}.{\bf 0012}9326 & {\bf  0}.{\bf 0012}6513 & {\bf  -0}.{\bf 0008}2044\\
 2 & {\bf  0}.{\bf 0020}2875 & {\bf  -0}.{\bf 0015}4161 & {\bf  0}.{\bf 00178}522 & {\bf  -0}.{\bf 0012}9216 & {\bf  0}.{\bf 0012}7530 & {\bf  -0}.{\bf 0008}1906 \\
 4 & {\bf  0}.{\bf 00204}843 & {\bf  -0}.{\bf 00156}428 & {\bf  0}.{\bf 00182}311 & {\bf  -0}.{\bf 0013}3177 & {\bf  0}.{\bf 0013}1916 & {\bf  -0}.{\bf 0008}5585 \\
 6 & {\bf  0}.{\bf 002049}19 & {\bf  -0}.{\bf 001565}16 & {\bf  0}.{\bf 00182}822 & {\bf  -0}.{\bf 00133}724 & {\bf  0}.{\bf 00132}841 & {\bf  -0}.{\bf 00086}363 \\
 8 & {\bf  0}.{\bf 0020492}1 & {\bf  -0}.{\bf 0015651}9 & {\bf  0}.{\bf 001829}09 & {\bf  -0}.{\bf 001338}18 & {\bf  0}.{\bf 00133}140 & {\bf  -0}.{\bf 00086}616 \\
 10 & & & {\bf  0}.{\bf 001829}37 & {\bf  -0}.{\bf 001338}33 & {\bf  0}.{\bf 001331}73 & {\bf  -0}.{\bf 000866}79 \\ 
%\bottomrule
\end{tabular}
\end{ruledtabular}
\end{table*}
%%%%%%%%%%%%%%%%%%%%%%%%%%%%

This technique is much less effective for the internuclear distances of atomic scale, because too much harmonics in the expansion~\eqref{eq:7} are required.  The energy of the $1\sigma_g$ level in the quasi-molecule $\rm U_2^{183+}$ with the distance $d=2/Z~\text{a.u.}\simeq 1150~\text{fm}$ is listed in Tab.~\ref{tab:eNum2Z} for the cutoff up to $\k_{max}<50$. The results of extrapolation into the region $\k_{max}<200$ coincide with the results of the Refs.~\cite{Kullie2001, Tupitsyn2010, Mironova2015, Artemyev2010} within the achievable accuracy of the 4 decimal digits. Thus, the method can be successfully applied for the internuclear distances up to $d \sim100$~fm, which is enough for describing the critical phenomena, however, for larger $d$ the convergence of the method turns out to be too slow.

%%%%%%%%%%%%%%%%%%%%%%%%%
\section{Results for the pure Coulomb problem}\label{sec:coulomb}
\begin{figure}[tbh]
\includegraphics[width=.48\textwidth]{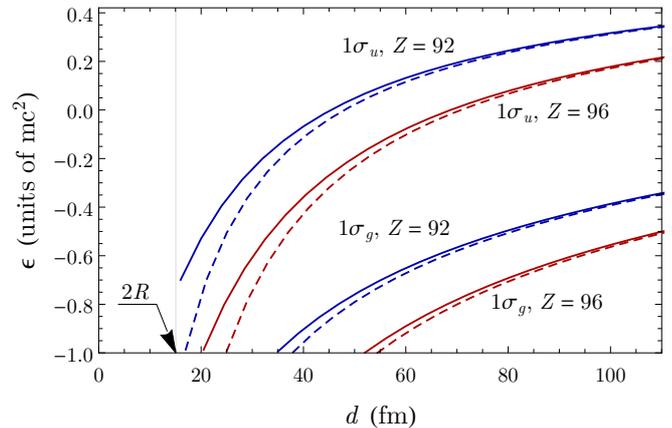}
\caption{(Color online) The electronic levels $1\sigma_g$ and $1\sigma_u$ in dependence on the internuclear distance $d=2a$ in two-nuclei quasi-molecules  $\rm U_2^{183+}$ (blue) and $\rm Cm_2^{191+}$ (red) in the pure Coulomb problem for the point-like (dashed) and extended (solid) nuclei.}\label{pic:Ed}\label{pic:Ed96}
\end{figure}

The energy of the lowest even $1\sigma_g$ and odd $1\sigma_u$ electronic levels in the compact two-nuclei quasi-molecules $\rm U_2^{183+}$ and $\rm Cm_2^{191+}$ in dependence on the internuclear distance $d$ for the cases of point-like and extended nuclei are plotted in the Fig.~\ref{pic:Ed}. For the system $\rm U-U$ the electronic level $1\sigma_g$ reaches the threshold of the lower continuum at $d=R_{cr}\simeq 34.75$~fm, while the lowest odd level $1\sigma_u$ dives only for the point-like nuclei and remains in the discrete spectrum for the minimal achieved distance between the extended nuclei (within the used nuclear charge distribution). For heavier nuclei (for example, $Z=96$ in Fig.~\ref{pic:Ed96}), starting from some distance, both levels dive into the negative continuum.

%%%%%%%%%%%%%%%%%%%%%%%%%
%
%	Fig.2
%
%								v3
%						(Color online)
%%%%%%%%%%%%%%%%%%%%%%%

The calculated critical distances $R_{cr}$ in the symmetric  quasi-molecules for the lowest even and odd electronic levels $1\sigma_g$ and $1\sigma_u$ are presented in Tab.~\ref{tab:Rcr} for the nuclear charge $Z\sim 87-100$, as well as the results from the Refs.~\cite{Mironova2015, Lisin1980, Marsman2011}. %For the nuclear charge $Z=85, 86$ the critical distances is less than the nuclear diameter.
The difference between these results is caused by using various nuclear charge distributions and nuclear radii. In the Ref.~\cite{Lisin1980} the calculations have been performed by the variational method, while the finite size of the nuclei has been taken into account in the quasi-classical approximation. The most precise computations~\cite{Mironova2015} have been carried out within the approach based on the two-center WF expansion by means of the experimental values for the nuclear radii. The results of $R_{cr}$ for the $1\sigma_g$ and $1\sigma_u$ electronic levels in the Refs.~\cite{Wietschorke1979, Soff1979} have been presented as a plot, hence, the corresponding values are not included in Tab.~\ref{tab:Rcr}. In the Ref.~\cite{Marsman2011} the critical distances have been calculated via the multipole expansion of the Coulomb potential only up to order $l_{max}=4$ and the same truncation $\kappa_{max}=4$ in the WF expansion, which, as it  follows from the Sec.~\ref{sec:pre} and Tabs.~\ref{tab:eNumA},~\ref{tab:eNumB}, is not enough to determine the energy of the electronic levels and $R_{cr}$ with high accuracy, especially for the most heavy nuclei in this region (see the third column in Tab.~\ref{tab:Rcr}).

%%%%%%%%%%%%%%%%%%%%%%%%%%%%%
%
%	Tab.4
%
%%%%%%%%%%%%%%%%%%%%%%%%%%%
\begin{table}[tbh]
%\center
\caption{The critical distances $R_{cr}$ in the symmetric two-nuclei quasi-molecules for the electronic levels $1\sigma_g$ and $1\sigma_u$ (in fm).}\label{tab:Rcr}
%\footnotesize
\begin{ruledtabular}
\begin{tabular}{i{3.0} i{2.2} i{3.17} i{2.2} }
%\toprule
\mc{ $Z$} & \mc{ $R_{cr} \ {(1\sigma_g)}$ }& \mc{ $R_{cr} \ (1\s_g\, , \, \text{other})$} & \mc{ $R_{cr} \ {(1\sigma_u)} $} \\
\colrule\noalign{\smallskip}
% 85 & & & & & 15.61^{a,b} & \\
% 86 & \text{---} & 12.86^a \quad 12.7^b & \\
% 86 & & & & & 18.29^{a,b} & \\
 87 & 16.20 & 16.42\footnotemark[1] \quad 16.0\footnotemark[2] \quad  & \\
 88 & 19.69 & 19.89\footnotemark[1] \quad 19.4\footnotemark[2] \quad 19.88\footnotemark[3] & \\
 89 & 23.27 & 23.38\footnotemark[1] \quad 22.9\footnotemark[2] \quad  & \\
 90 & 26.96 & 26.96\footnotemark[1] \quad 26.5\footnotemark[2] \quad 26.88\footnotemark[3] & \\
 91 & 30.78 & 30.90\footnotemark[1] \quad 30.3\footnotemark[2] \quad  &\\
 92 & 34.75 & 34.72\footnotemark[1] \quad 34.3\footnotemark[2] \quad 34.38\footnotemark[3] &\\
 93 & 38.85 & 38.93\footnotemark[1] \quad 38.4\footnotemark[2] \quad  &\\
% \cmidrule{1-2}
 94 & 43.10 & 43.10\footnotemark[1] \quad 42.6\footnotemark[2] \quad 42.52\footnotemark[3] & 15.42  \\
 95 & 47.49 & 47.47\footnotemark[1] \quad 47.0\footnotemark[2] \quad  & 17.82  \\
 96 & 52.01 & 52.06\footnotemark[1] \quad 51.6\footnotemark[2] \quad 51.07\footnotemark[3] & 20.25  \\
 97 & 56.68 & 56.77\footnotemark[1] \quad 56.3\footnotemark[2] \quad  & 22.73  \\
 98 & 61.48 & 61.56\footnotemark[1] \quad 61.0\footnotemark[2] \quad 60.08\footnotemark[3] & 25.26  \\
 99 & 66.41 & 66.50\footnotemark[1] \quad 66.0\footnotemark[2] \quad  & 27.86  \\
 100 & 71.46 & 71.57\footnotemark[1] \quad 71.1\footnotemark[2] \quad  & 30.53  \\
%\botrule
\end{tabular}
\end{ruledtabular}
\footnotetext[1]{From Ref.~\cite{Mironova2015}}
\footnotetext[2]{From Ref.~\cite{Lisin1980}}
\footnotetext[3]{From Ref.~\cite{Marsman2011}}
%\begin{flushleft}
%\hspace{0em}${}^a$ \text{Ref.~\cite{Mironova2015}}, ${}^b$ \text{Ref.~\cite{Lisin1980}}, ${}^c$ \text{Ref.~\cite{Marsman2011}}.
%\end{flushleft}
\end{table}
%%%%%%%%%%%%%%%%%%%%%%%%

%%%%%%%%%%%%%%%%%%%%%%%%%%
\section{The levels shift due to $\Delta U_{AMM}$}\label{sec:nz}

By comparing the solutions of Eqs.~\eqref{eq:8} with the pure Coulomb case (Eqs.~\eqref{eq:8} with $\l=0$), one obtains the shift of the electronic levels in a two-nuclei quasi-molecule due to $\Delta U_{AMM}$. This shift is calculated completely non-perturbatively in $Z\alpha$ and (partially) in $\a/\pi$, since the latter enters as a factor in the coupling constant for $\Delta U_{AMM}$. It should be underlined that this kind of nonlinearity in $\a/\pi$ has nothing to do with the  summation of the loop expansion for AMM, since the initial expression for the operator~\eqref{eq:04} is based on the one-loop approximation for the vertex \cite{Roenko2017}. The shifts of the $1\sigma_g$ and $1\sigma_u$ levels are presented in Tab.~\ref{tab:dEamm} for various nuclear charges $Z$, when the internuclear distance $d$ is fixed. The empty cells in the Tab.~\ref{tab:dEamm} mean that for the corresponding values of  $Z$ and $d$ the level has already sunk into the lower continuum. The shift of the lowest even level turns out to be positive and the shift of the lower odd one is negative; their absolute magnitude is about $1~\text{keV}$, when the electronic level lies near the lower threshold with $\epsilon \simeq -1$ (i.e. when $Z\simeq Z_{cr}$), as in the case of an H-like atom (see Ref.~\cite{Roenko2017}).
It should be mentioned that the calculations of the total self-energy shift of the $1s_{1/2}$ level for an H-like atom with nuclear charge $Z_{cr}\simeq 170$, performed in Refs.~\cite{Cheng1976,Soff1982} to the leading order in $\a/\pi$ with complete dependence on $Z\a$, give  $\D E_{SE}(1s_{1/2}) \hm\simeq 11.0 \text{\ keV} $. So the shift due to AMM in H-like atoms turns out to be about the tenth part of the self-energy contribution to the full radiative shift (for $1s_{1/2}$) and its magnitude is comparable with the higher-order corrections to vacuum polarization effects~\cite{Gyulassy1974,Gyulassy1975}. However, the most important point here is not the magnitude, but the observation that the behavior of the shift due to AMM qualitatively reproduces the behavior of the whole self-energy contribution for the  lowest electronic levels~\cite{Roenko2017, Roenko2017a, Roenko2017b}.

%For comparison, like to AMM, the higher-order vacuum polarization effects make a similar contribution of about 10\% ($\Delta E_{VP}^{(3+)}(1s_{1/2})\hm\simeq-1.1$~keV) to the total vacuum polarization shift of the level $\Delta E_{VP}(1s_{1/2})\hm\simeq-10.7$~keV~\cite{Pieper1969,Gyulassy1974,Gyulassy1975}.

%For the H-like atom with nuclear charge $Z_{cr}=170$ calculations, performed in Refs.~\cite{Cheng1976, Soff1982} to the leading order in $\a/\pi$ with complete dependence on $Z\a$, give  $\D E_{SE}(1s_{1/2}) \hm\simeq 11.0 \text{\ keV} $

%%%%%%%%%%%%%%%%%%%%%%%%%%
%
%	Tab. 5
%
%%%%%%%%%%%%%%%%%%%%%%%%%%
\begin{table*}[tbh]
%\center
\caption{The shift of the levels $1\sigma_g$ and $1\sigma_u$ due to $\Delta U_{AMM}$ for a various total charge of nuclei $Z_{\Sigma}=2Z$ and internuclear distances $d$. The shift of the levels $1s_{1/2}$ and $2p_{1/2}$ in H-like atom from Ref.~\cite{Roenko2017} is shown for comparison (in keV).}\label{tab:dEamm}
\begin{ruledtabular}
\begin{tabular}{c c *{5}{i{2.3}}}
%\toprule
\mc{ level } & \mc{ $Z_{\Sigma}$ } & \mc{ H-like atom } & \mc{ $d=15.5$~fm } & \mc{ $d=20$~fm } & \mc{ $d=30$~fm } & \mc{ $d=40$~fm }\\
\colrule\noalign{\smallskip}
\multirow{8}{*}{\minitab[c]{ $1\sigma_g$ \\ ($1s_{1/2}$)} } & 140 & 0.495 & 0.465 & 0.448 & 0.413 & 0.385 \\
 & 150 & 0.690 & 0.635 & 0.603 & 0.545 & 0.500 \\
 & 160 & 0.912 & 0.828 & 0.779 & 0.692 & 0.626 \\
 & 170 & 1.118 & 1.017 & 0.953 & 0.840 & 0.755 \\
 & 173 & \mc{ --- } & 1.068 & 1.002 & 0.883 & 0.793 \\
 & 176 & \mc{ --- } & \mc{ --- } & 1.047 & 0.924 & 0.830 \\
 & 181 & \mc{ --- } & \mc{ --- } & \mc{ --- } & 0.987 & 0.888 \\
 & 186 & \mc{ --- } & \mc{ --- } & \mc{ --- } & \mc{ --- } & 0.942 \\
\colrule\noalign{\smallskip}
\multirow{10}{*}{\minitab[c]{ $1\sigma_u$ \\ ($2p_{1/2}$)} } & 150 & -0.373 & -0.329 & -0.304 & -0.264 & -0.234 \\
 & 160 & -0.632 & -0.546 & -0.497 & -0.417 & -0.361 \\
 & 170 & -0.875 & -0.763 & -0.696 & -0.580 & -0.498 \\
 & 180 & -1.052 & -0.937 & -0.861 & -0.725 & -0.625 \\
 & 183 & -1.090 & -0.978 & -0.901 & -0.763 & -0.659 \\
 & 188 & \mc{ --- } & -1.034 & -0.960 & -0.819 & -0.711 \\
 & 191 & \mc{ --- } & \mc{ --- } & -0.989 & -0.848 & -0.738 \\
 & 195 & \mc{ --- } & \mc{ --- } & \mc{ --- } & -0.883 & -0.773 \\
 & 199 & \mc{ --- } & \mc{ --- } & \mc{ --- } & -0.912 & -0.802 \\
 & 206 & \mc{ --- } & \mc{ --- } & \mc{ --- } & \mc{ --- } & -0.843 \\
%\botrule
\end{tabular}
\end{ruledtabular}
\end{table*}
Another valuable characteristics of QED-effects is their  growth rate as a function of Z. The shift due to $\Delta U_{AMM}$ for an atomic electron is a part of the self-energy contribution to the Lamb shift, which in the perturbative QED is proportional to $Z^4/n^3$ and usually represented through the function $F_{nj}(Z\a)$, defined via~\cite{Mohr1998}
\begin{equation} \label{eq:21}
\Delta E^{SE}_{nj}(Z\alpha)=\frac{ Z^4 \alpha^5}{\pi n^3}\, F_{nj}(Z\alpha)  \, .
\end{equation}
The function $F_{nj}(Z\alpha)$ is approximated via known data for a number of $Z \lesssim 100$ as a slowly varying function of $Z\alpha$~\cite{Greiner2003, Mohr1998, Pyykko2012, Johnson1985, Yerokhin2015}. The calculation of the atomic levels shift, stipulated by the Dirac-Pauli term, has been preformed in Refs.~\cite{Barut1982, Sveshnikov2013, Sveshnikov2016}, but in the case of strong fields this approximation is too rough, and accounting for the dependence of $F_2(q^2)$ on the momentum transfer (as in Eq.~\eqref{eq:06}) in this case is crucial~\cite{Roenko2017}.

\begin{figure*}[tbh]
\subfigure{\label{pic:2N-1s-F}
\includegraphics[width=.48\textwidth]{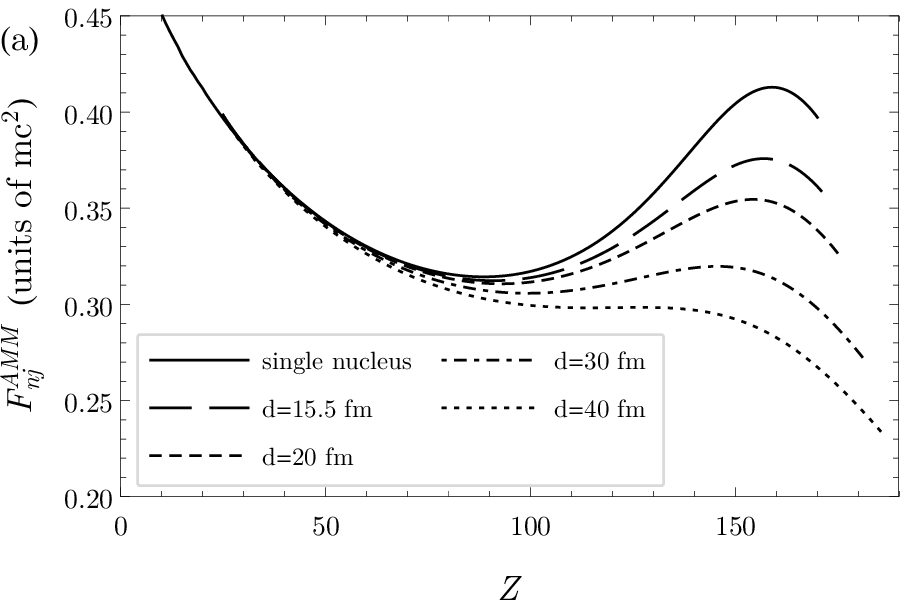}
}
\hfill
\subfigure{\label{pic:2N-2p-F}
\includegraphics[width=.48\textwidth]{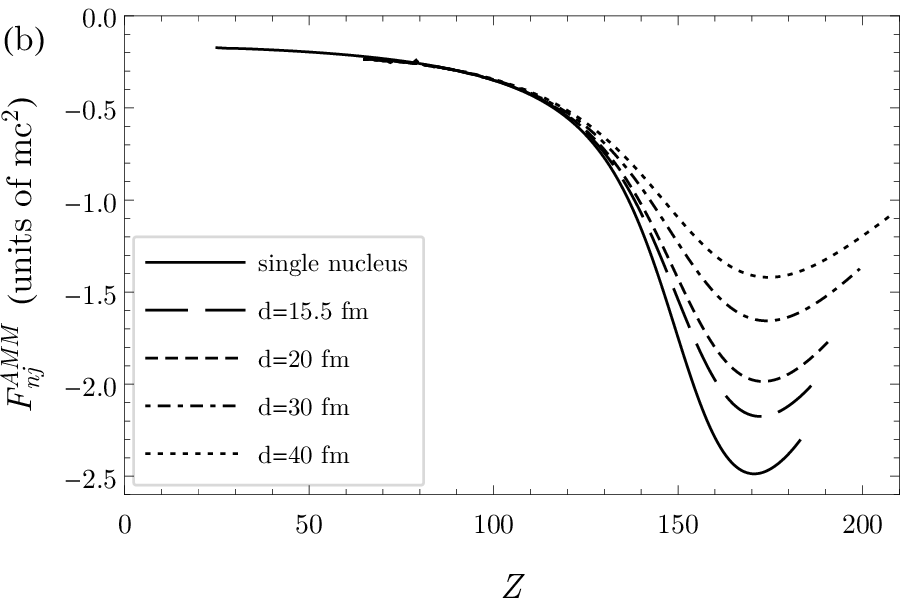}
}
\subfigure{\label{pic:2N-1s-n}
\includegraphics[width=.48\textwidth]{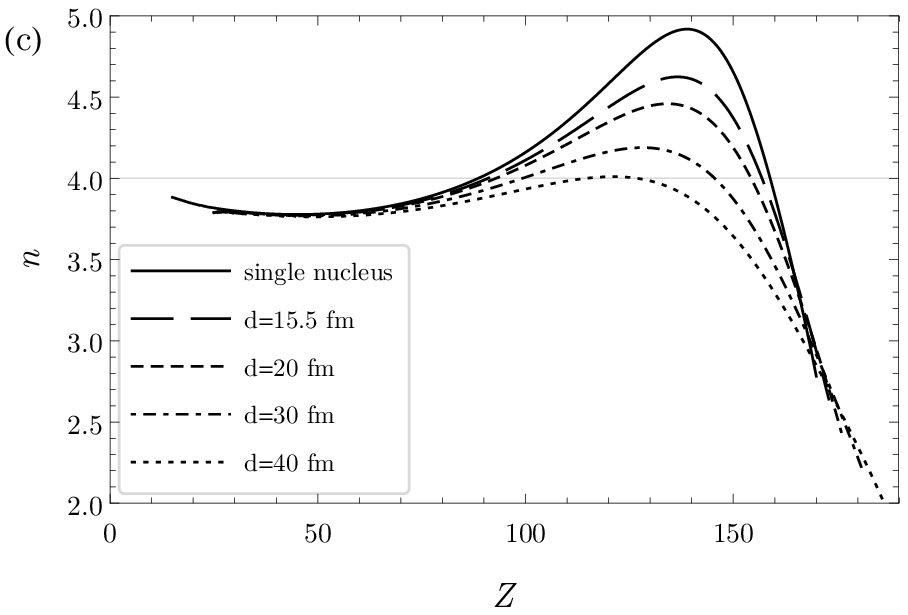}
}
\hfill
\subfigure{\label{pic:2N-2p-n}
\includegraphics[width=.48\textwidth]{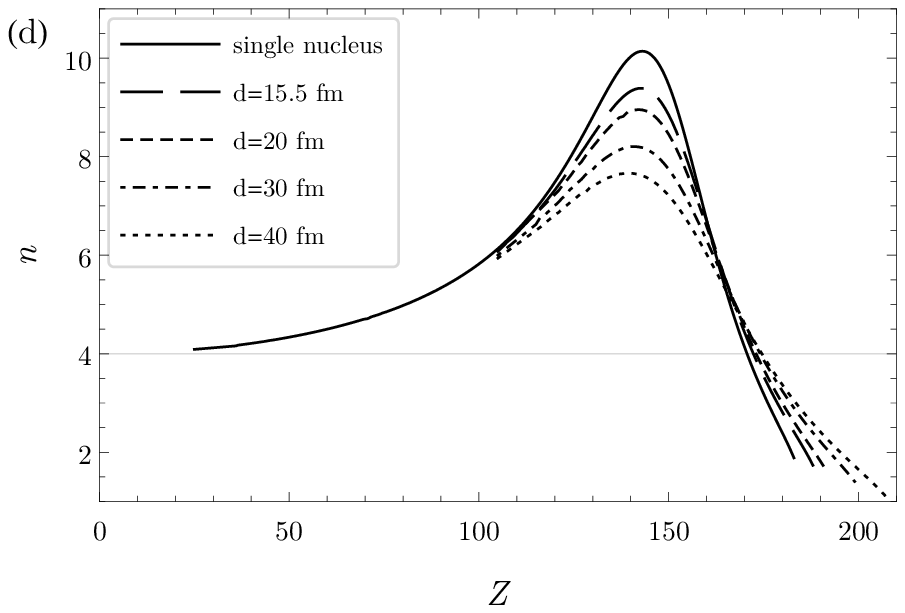}
}
\caption{The function $F_{nj}^{AMM}$ and the rate of growth $n(Z)$ for the shift due to $\D U_{AMM}$ of the $1\sigma_g$~\subref{pic:2N-1s-F},~\subref{pic:2N-1s-n} and $1\sigma_u$~\subref{pic:2N-2p-F},~\subref{pic:2N-2p-n} levels in a symmetrical two-nuclei quasi-molecule as a function of the total charge $Z$ for the fixed internuclear distances $d=15.5, 20, 30, 40$~fm. For comparison, the same functions for the $1s_{1/2}$ and $2p_{1/2}$ levels in H-like atom from Ref.~\cite{Roenko2017} are shown.}\label{pic:2N-F}
\end{figure*}

It would be convenient to compare the shift of the electronic levels in a very compact quasi-molecule with the shift for a single nuclei with $Z\simeq Z_{cr}$. The function $F_{nj}^{AMM}(Z_\Sigma \alpha)$ (where $Z_\Sigma$ is the total charge of the Coulomb sources) for the lowest even and odd levels in the quasi-molecule for a range of internuclear distances $d$ and large $Z$ is plotted in Figs.~\ref{pic:2N-1s-F},~\ref{pic:2N-2p-F}, combined with the case of an H-like atom from Ref.~\cite{Roenko2017}. It follows from the Figs.~\ref{pic:2N-1s-F},~\ref{pic:2N-2p-F} and Tab.~\ref{tab:dEamm} that in the quasi-molecule  the  shift due to $\D U_{AMM}$ decreases with the growing  internuclear distance quite rapidly. Moreover, the shift of the electronic levels near the threshold of the lower continuum also falls down with the increasing critical charge $Z_{cr}$, as well as with the corresponding size of the quasi-molecule $R_{cr}$ (see the lowest values in the each column in Tab.~\ref{tab:dEamm}).

%%%%%%%%%%%%%%%%%%%%%%%%%%%%
%
%	Fig. 3
%
%%%%%%%%%%%%%%%%%%%%%%%%%%%%

In Figs.~\ref{pic:2N-1s-n},~\ref{pic:2N-2p-n}, the power-like approximation $n(Z_\Sigma)$  of the growth rate for the shift of  $1\sigma_g$ and $1\sigma_u$ levels in the quasi-molecule is shown as a function of the total charge $Z_\Sigma$, determined via logarithmic derivative
\begin{equation} \label{eq:22}
n(Z_\Sigma)= Z_\Sigma \frac{  \partial }{ \partial Z_\Sigma} \ln \( \Delta \e_{AMM} \)\, .
\end{equation}
All the curves in Fig.~\ref{pic:2N-F} for the quasi-molecule become closer and closer to the case of a single nuclei, when the internuclear distance tends to zero. The growth rates in Fig.~\ref{pic:2N-F} reveal their maximums for $Z\sim 140-150$, while their magnitude  decreases rapidly with the increasing  size of the quasi-molecules. However, the further decline in $n(Z)$ in the region $Z > 150$ remains still pronounced. At the same time, for $d=40$~fm  the function $F_{1\sigma g}^{AMM}(Z_\Sigma \alpha)$ shows up already an almost monotonic behavior.

The functions, that are shown in the Fig.~\ref{pic:2N-F},  depend weakly on the  truncation used in the expansion~\eqref{eq:7}, and so look like in the monopole approximation, when only the spherical-symmetric part of the two-center potential is taken into account, which corresponds to the charge distribution over the sphere with diameter $d$. This circumstance does not contradict to the fact that the multipole moment $V_0$ is much bigger than the other multipoles $V_n$, as it has been  noted in Sec.~\ref{sec:formfact}. At the same time, for the general properties of the levels shift due to AMM, the size of the source, which produces the critical field, is much more important, than the particular model of the charge distribution.

\section{Conclusion}\label{sec:concl}
To conclude we have shown that the approach, based on  the multipole expansions~\eqref{eq:7},~\eqref{eq:11} for  solving the two-center DE, can be successfully applied for the compact nuclear quasi-molecules ($d\lesssim 100$~fm), and allows  to investigate the process of diving of discrete levels into the lower continuum in heavy ions collisions not only qualitatively (what could be performed within the monopole approximation), but also quantitatively. Calculated critical distances $R_{cr}$ for the levels $1\sigma_g$ and $1\sigma_u$ are in a good agreement with other computations~\cite{Mironova2015, Lisin1980, Marsman2011, Tupitsyn2010, Rafelski1976, Soff1979, Wietschorke1979}. Moreover, the results $R_{cr}$ for the $1\sigma_u$ level significantly improve the most relevant values, which have been obtained early within the monopole approximation~\cite{Wietschorke1979}.

   By means of the same techniques the effective interaction of the electronic AMM with the Coulomb potential of colliding nuclei is considered with account of the dynamical screening of $\D U_{AMM}$ at small distances due to dependence of the electronic form factor $F_2(q^2)$ on the momentum transfer. The performed study of the electronic levels in a compact quasi-molecule due to $\Delta U_{AMM}$ as a function of the nuclear charge and internuclear distance shows that the shift of levels near the lower continuum decreases with the increasing $Z_\Sigma$ and the distance between nuclei (that means with enlarging the size of the system of Coulomb sources) both in absolute units and in terms of $Z_\Sigma^4 \a^5 / \pi n^3$. The rate of growth of this QED-effect, defined via~\eqref{eq:22}, behaves more smoothly with the increasing  internuclear distance, and shows up a stable decline in the region $Z_\Sigma >150$.

And although the shift due to AMM  is just a part of the whole radiative correction to the binding energy, 
    the behavior of $F_{nj}^{AMM}(Z\alpha)$ qualitatively reproduces the behavior of $F_{nj}(Z\alpha)$ for the  lowest electronic levels~\cite{Roenko2017, Roenko2017a, Roenko2017b}. Thus, there appears a natural assumption, that in the overcritical region the decrease with the growing  $Z_\Sigma$  and the size  of the system of Coulomb sources should take place also for the total self-energy contribution to  the levels shift near the threshold of the lower continuum, and so for the other radiative QED-effects with virtual photon exchange.

\begin{acknowledgments}
The authors are very indebted to Prof.~P.~K.~Silaev from MSU Department of Physics for interest and helpful discussions.
This work has been supported in part by the RF Ministry of Education and Science Scientific Research Program, Projects No. 01-2014-63889 and No. A16-116021760047-5, and by RFBR Grant No. 14-02-01261.
\end{acknowledgments}

%%%%%%%%%%%%%%%%%%%%%%%%%%%
\onecolumngrid

\appendix
\renewcommand{\theequation}{\thesection.\arabic{equation}}                %    меняем номер формул на "номер раздела"."номер формулы"
%\begin{widetext}
%%%%%%%%%%%%%%%%%%%%%%%%%
%\begin{widetext}
\section{The system of equations for the levels of definite parity}\label{sec:app-syst}
The expansion in spherical spinors~\eqref{eq:7} and the system of equations~\eqref{eq:8} for the even level with definite $m_j$ take the form~\eqref{eq:apev} (the notations $u_k= f_{-2k-1}$, $v_k= f_{2k+2}$, $p_k= g_{2k+2}$, $q_k= g_{-2k-1}$ are used):
\begin{subequations}\label{eq:apev}%
\begin{equation}\label{eq:7evvv}
\varphi = \sum_{k=0}^{\tilde{N}} \left( u_k \, X_{-2k-1,m_j} + v_k \, X_{2k+2,m_j} \right) , \quad \chi =\sum_{k=0}^{\tilde{N}} \left( p_k \, X_{-2k-2,m_j} + q_k \, X_{2k+1,m_j} \right),
\end{equation}
\begin{equation}\label{eq:8evvv}
\begin{split}
	\partial_r u_{k} &- \frac{2k}{r}u_{k}  + \lambda \sum_{\bar{k}} \Big( M_{-2k-1;-2\bar{k}-1}(r)\, u_{\bar{k}} +  M_{-2k-1;2\bar{k}+2}(r)\, v_{\bar{k}} \Big) = \\
	& \hspace{12em} =(1 + \epsilon) q_{k} + \alpha \sum_{\bar{k}} \Big(  N_{2k+1;2\bar{k}+1}(r)\, q_{\bar{k}} + N_{2k+1;-2\bar{k}-2}(r)\, p_{\bar{k}} \Big),\\
	\partial_r v_{k} &+ \frac{2k+3}{r}v_{k} + \lambda \sum_{\bar{k}} \Big( M_{2k+2;-2\bar{k}-1}(r)\, u_{\bar{k}} +  M_{2k+2;2\bar{k}+2}(r)\, v_{\bar{k}} \Big) = \\
	& \hspace{12em} = (1 + \epsilon) p_{k} + \alpha \sum_{\bar{k}} \Big(  N_{-2k-2;2\bar{k}+1}(r)\, q_{\bar{k}} + N_{-2k-2;-2\bar{k}-2}(r)\, p_{\bar{k}} \Big),\\
	\partial_r p_{k} &- \frac{2k+1}{r}p_{k} - \lambda \sum_{\bar{k}} \Big( M_{-2k-2;2\bar{k}+1}(r)\, q_{\bar{k}} + M_{-2k-2;-2\bar{k}-2}(r)\, p_{\bar{k}} \Big) = \\
	& \hspace{12em} = (1 - \epsilon)v_{k} - \alpha \sum_{\bar{k}}  \Big( N_{2k+2;-2\bar{k}-1}(r)\, u_{\bar{k}} +  N_{2k+2;2\bar{k}+2}(r)\, v_{\bar{k}} \Big),\\
	\partial_r q_{k} &+ \frac{2k+2}{r}q_{k} - \lambda \sum_{\bar{k}} \Big(  M_{2k+1;2\bar{k}+1}(r)\, q_{\bar{k}} + M_{2k+1;-2\bar{k}-2}(r)\, p_{\bar{k}}  \Big) = \\
	& \hspace{12em} =(1 - \epsilon)u_{k} - \alpha \sum_{\bar{k}} \Big( N_{-2k-1;-2\bar{k}-1}(r)\, u_{\bar{k}} +  N_{-2k-1;2\bar{k}+2}(r)\, v_{\bar{k}} \Big),\\
\end{split}
\end{equation}
\end{subequations}
and for odd ---~\eqref{eq:apod} (the notations $u_k= f_{-2k-2}$, $v_k= f_{2k+1}$, $p_k= g_{2k+1}$, $q_k= g_{-2k-2}$ are used):
\begin{subequations}\label{eq:apod}%
\begin{equation}\label{eq:7oddd}
\varphi = \sum_{k=0}^{\tilde{N}} \left( u_k \, X_{-2k-2,m_j} + v_k \, X_{2k+1,m_j} \right) , \quad \chi =\sum_{k=0}^{\tilde{N}} \left( p_k \, X_{-2k-1,m_j} + q_k \, X_{2k+2,m_j} \right),
\end{equation}
\begin{equation}\label{eq:8oddd}
\begin{split}
	\partial_r u_{k} &- \frac{2k+1}{r}u_{k}  + \lambda \sum_{\bar{k}} \Big( M_{-2k-2;-2\bar{k}-2}(r)\, u_{\bar{k}} +  M_{-2k-2;2\bar{k}+1}(r)\, v_{\bar{k}} \Big) = \\
	& \hspace{12em} =(1 + \epsilon) q_{k} + \alpha \sum_{\bar{k}} \Big(  N_{2k+2;2\bar{k}+2}(r)\, q_{\bar{k}} + N_{2k+2;-2\bar{k}-1}(r)\, p_{\bar{k}} \Big),\\
	\partial_r v_{k} &+ \frac{2k+2}{r}v_{k} + \lambda \sum_{\bar{k}} \Big( M_{2k+1;-2\bar{k}-2}(r)\, u_{\bar{k}} +  M_{2k+1;2\bar{k}+1}(r)\, v_{\bar{k}} \Big) = \\
	& \hspace{12em} = (1 + \epsilon) p_{k}  + \alpha \sum_{\bar{k}} \Big(  N_{-2k-1;2\bar{k}+2}(r)\, q_{\bar{k}} + N_{-2k-1;-2\bar{k}-1}(r)\, p_{\bar{k}} \Big),\\
	\partial_r p_{k} &- \frac{2k}{r}p_{k} - \lambda \sum_{\bar{k}} \Big( M_{-2k-1;2\bar{k}+2}(r)\, q_{\bar{k}} + M_{-2k-1;-2\bar{k}-1}(r)\, p_{\bar{k}} \Big) = \\
	& \hspace{12em} = (1 - \epsilon)v_{k} - \alpha \sum_{\bar{k}}  \Big( N_{2k+1;-2\bar{k}-2}(r)\, u_{\bar{k}} +  N_{2k+1;2\bar{k}+1}(r)\, v_{\bar{k}} \Big),\\
	\partial_r q_{k} &+ \frac{2k+3}{r}q_{k} - \lambda \sum_{\bar{k}} \Big(  M_{2k+2;2\bar{k}+2}(r)\, q_{\bar{k}} + M_{2k+2;-2\bar{k}-1}(r)\, p_{\bar{k}}  \Big) = \\
	& \hspace{12em} = (1 - \epsilon)u_{k} - \alpha \sum_{\bar{k}} \Big( N_{-2k-2;-2\bar{k}-2}(r)\, u_{\bar{k}} +  N_{-2k-2;2\bar{k}+1}(r)\, v_{\bar{k}} \Big).\\
\end{split}
\end{equation}
\end{subequations}
%\end{widetext}

%%%%%%%%%%%%%%%%%%%%%%%%%%%%%%%%%%
\section{Analytic expressions of the $U_n$}\label{sec:app-Fn}
%\begin{widetext}
In the intermediate region $|r-a| \leq R$ the multipole moments~\eqref{eq:20} are given by the following analytic expressions (for $n \leq 12$):
\begin{flalign}
&\hspace{\parindent}U_0(r)=\frac{2Z}{16 a r R^3}\Big(a^4-4 a^3 r-4 a (r-2 R) (r+R)^2+(r-R)^3 (r+3 R)+6 a^2 \left(r^2-R^2\right)\Big). &
\end{flalign}
\begin{multline}
U_2 (r)=\frac{2Z}{256 a^3 r^3 R^3}\Big(5 a^8+128 a^5 R^3-20 a^2 \left(r^2-R^2\right)^3-20 a^6 \left(r^2+3 R^2\right)+{}\\
{}+(r-R)^5 \left(5 r^3+25 r^2 R+15 r R^2+3 R^3\right)+30 a^4 \left(r^4+2 r^2 R^2-3 R^4\right)\Big)
\end{multline}
\begin{multline}
U_4 (r)=\frac{2Z}{2048 a^5 r^5 R^3}\Big(21 a^{12}+1024 a^9 R^3-54 a^2 \left(r^2-R^2\right)^5-54 a^{10} \left(r^2+7 R^2\right)+ 27 a^4 \left(r^2-R^2\right)^3 \left(r^2+7 R^2\right)+{}\\
{}+27 a^8 \left(r^4+10 r^2 R^2-35 R^4\right)+   (r-R)^7 \left(21 r^5+147 r^4 R+210 r^3 R^2+142 r^2 R^3+49 r R^4+7 R^5\right)+{} \\
+12 a^6 \left(r^6+9 r^4 R^2-45 r^2 R^4+35 R^6\right)\Big)
\end{multline}
\begin{multline}
U_6 (r)=\frac{2Z}{65536 a^7 r^7 R^3}\Big(429 a^{16}+32768 a^{13} R^3-936 a^2 \left(r^2-R^2\right)^7-936 a^{14} \left(r^2+11 R^2\right)+{}\\ {}+364 a^4 \left(r^2-R^2\right)^5 \left(r^2+11 R^2\right)+ 364 a^{12} \left(r^4+18 r^2 R^2-99 R^4\right)+{}\\ {}+104 a^6 \left(r^2-R^2\right)^3 \left(r^4+18 r^2 R^2-99 R^4\right)+ 104 a^{10} \left(r^6+21 r^4 R^2-189 r^2 R^4+231 R^6\right)+{}\\ {}+(r-R)^9 \big(429 r^7+3861 r^6 R+9009 r^5 R^2+10889 r^4 R^3+7911 r^3 R^4+3519 r^2 R^5+891 r R^6+99 R^7\big)+{}\\ {}+78 a^8 \left(r^8+20 r^6 R^2-210 r^4 R^4+420 r^2 R^6-231 R^8\right)\Big)
\end{multline}
\begin{multline}
 U_8 (r)=\frac{2Z}{524288 a^9 r^9 R^3}\Big(2431 a^{20}+262144 a^{17} R^3-4862 a^2 \left(r^2-R^2\right)^9-4862 a^{18} \left(r^2+15 R^2\right)+{}\\ {}+1683 a^4 \left(r^2-R^2\right)^7 \left(r^2+15 R^2\right)+1683 a^{16} \left(r^4+26 r^2 R^2-195 R^4\right)+{}\\ {}+408 a^6 \left(r^2-R^2\right)^5 \left(r^4+26 r^2 R^2-195 R^4\right)+408 a^{14} \left(r^6+33 r^4 R^2-429 r^2 R^4+715 R^6\right)+{}\\ {}+238 d^8 \left(r^2-R^2\right)^3 \left(r^6+33 r^4 R^2-429 r^2 R^4+715 R^6\right)   +238 a^{12} \left(r^8+36 r^6 R^2-594 r^4 R^4+1716 r^2 R^6-1287 R^8\right)+{}\\ {}+(r-R)^{11} \big(2431 r^9+26741 r^8 R+87516 r^7 R^2+155180 r^6 R^3+175450 r^5 R^4+133782 r^4 R^5+ 69212 r^3 R^6+{}\\+23452 r^2 R^7+4719 r R^8+429 R^9\big) {}+204 a^{10} \left(r^{10}+35 r^8 R^2-630 r^6 R^4+2310 r^4 R^6-3003 r^2 R^8+1287 R^{10}\right)\Big)
\end{multline}
\begin{multline}
U_{10} (r)=\frac{2Z}{8388608 a^{11} r^{11} R^3}\Big(29393 a^{24}+4194304 a^{21} R^3-55692 a^2 \left(r^2-R^2\right)^{11}-{}\\ {}-55692 a^{22} \left(r^2+19 R^2\right)+18018 a^4 \left(r^2-R^2\right)^9 \left(r^2+19 R^2\right)+18018 a^{20} \left(r^4+34 r^2 R^2-323 R^4\right)+{}\\
{}+4004 a^6 \left(r^2-R^2\right)^7 \left(r^4+34 r^2 R^2-323 R^4\right)+4004 a^{18} \left(r^6+45 r^4 R^2-765 r^2 R^4+1615 R^6\right)+{}\\
{}+2079 a^8 \left(r^2-R^2\right)^5 \left(r^6+45 r^4 R^2-765 r^2 R^4+1615 R^6\right)+2079 a^{16} \big(r^8+52 r^6 R^2-1170 r^4 R^4+4420 r^2 R^6-{}\\
{}-4199 R^8\big)+1512 a^{10} \left(r^2-R^2\right)^3 \left(r^8+52 r^6 R^2-1170 r^4 R^4+4420 r^2 R^6-4199 R^8\right)+{}\\
{}+216 a^{14} \left(7 r^{10}+385 r^8 R^2-10010 r^6 R^4+50050 r^4 R^6-85085 r^2 R^8+46189 R^{10}\right)+{}\\
{}+(r-R)^{13} \big(29393 r^{11}+382109 r^{10} R+1616615 r^9 R^2+3812195 r^8 R^3+5909930 r^7 R^4+\\+6450626 r^6 R^5+5093998 r^5 R^6+2916550 r^4 R^7+1186549 r^3 R^8+326417 r^2 R^9+54587 r R^{10}+4199 R^{11}\big)+\\+196 a^{12} \big(7 r^{12}+378 r^{10} R^2-10395 r^8 R^4+60060 r^6 R^6-135135 r^4 R^8+131274 r^2 R^{10}-46189 R^{12}\big)\Big)
\end{multline}
\begin{multline}
U_{12} (r)=
\frac{2Z}{67108864 a^{13} r^{13} R ^3}\Big(185725 a^{28}+33554432 a^{25} R ^3-339150 a^2 \left(r^2-R ^2\right)^{13}-339150 a^{26} \left(r^2+23 R ^2\right)+{}\\
{}+104975 a^4 \left(r^2-R ^2\right)^{11} \left(r^2+23 R ^2\right)+104975 a^{24} \left(r^4+42 r^2 R ^2-483 R ^4\right)+{}\\
{}+22100 a^6 \left(r^2-R ^2\right)^9 \left(r^4+42 r^2 R ^2-483 R ^4\right)+22100 a^{22} \left(r^6+57 r^4 R ^2-1197 r^2 R ^4+3059 R ^6\right)+{}\\
{}+10725 a^8 \left(r^2-R ^2\right)^7 \left(r^6+57 r^4 R ^2-1197 r^2 R ^4+3059 R ^6\right)+2145 a^{20} \big(5 r^8+340 r^6 R ^2-9690 r^4 R ^4+45220 r^2 R ^6-{}\\
{}-52003 R ^8\big)+1430 a^{10} \left(r^2-R ^2\right)^5 \left(5 r^8+340 r^6 R ^2-9690 r^4 R ^4+45220 r^2 R ^6-52003 R ^8\right)+{}\\
{}+7150 a^{18} \left(r^{10}+75 r^8 R ^2-2550 r^6 R ^4+16150 r^4 R ^6-33915 r^2 R ^8+22287 R ^{10}\right)+{}\\
{}+5775 a^{12} \left(r^2-R ^2\right)^3 \left(r^{10}+75 r^8 R ^2-2550 r^6 R ^4+16150 r^4 R ^6-33915 r^2 R ^8+22287 R ^{10}\right)+{}\\
{}+1925 a^{16} \left(3 r^{12}+234 r^{10} R ^2-8775 r^8 R ^4+66300 r^6 R ^6-188955 r^4 R ^8+226746 r^2 R ^{10}-96577 R ^{12}\right)+{}\\
{}+(r-R )^{15} \big(185725 r^{13}+2785875 r^{12} R +14486550 r^{11} R ^2+42840682 r^{10} R ^3+84878055 r^9 R ^4+121292265 r^8 R ^5+{}\\
{}+129590660 r^7 R ^6+105101820 r^6 R ^7+64747611 r^5 R ^8+29914645 r^4 R ^9+10067910 r^3 R ^{10}+2335290 r^2 R ^{11}+{}\\
{}+334305 r R ^{12}+22287 R ^{13}\big)+1800 a^{14} \big(3 r^{14}+231 r^{12} R ^2-9009 r^{10} R ^4+75075 r^8 R ^6-255255 r^6 R ^8+{}\\
{}+415701 r^4 R ^{10}-323323 r^2 R ^{12}+96577 R ^{14}\big)\Big)
\end{multline}

%\vspace{-1ex}
%%%%%%%%%%%%%%%%%%%%%%

\twocolumngrid

\bibliography{biblio/AMM}
\end{document}